\documentclass[usenatbib]{mn2e}
\usepackage{psfig,times}

\title[Magnetic activity and differential rotation on AB Dor]{Magnetic
activity on AB Doradus: Temporal evolution of starspots and
differential rotation from 1988 to 1994}

\author[S.V. Jeffers, J.-F. Donati, A.Collier Cameron]
        {S.V. Jeffers$^{1,2}$, J.-F. Donati $^1$, A.Collier
        Cameron$^{2}$\\ $^1$Laboratoire d'Astrophysique Toulouse-Tarbes, Observatoire
        Midi-Pyr$\acute{e}$n$\acute{e}$es, 14, avenue Edouard Belin,
        F-31400 Toulouse, France\\ $^2$School of Physics and
        Astronomy, University of St\ Andrews, North Haugh, St\
        Andrews, Fife KY16 9SS, UK \\ }

\date{}
        
\pagerange{\pageref{firstpage}--\pageref{lastpage}}
\pubyear{2006}

\begin{document}

\maketitle

\label{firstpage}

\begin{abstract}
Surface brightness maps for the young K0 dwarf AB Doradus are
reconstructed from archival data sets for epochs spanning 1988 to
1994.  By using the signal-to-noise enhancement technique of
Least-Squares Deconvolution, our results show a greatly increased
resolution of spot features than obtained in previously published
surface brightness reconstructions.  These images show that for the
exception of epoch 1988.96, the starspot distributions are dominated
by a long-lived polar cap, and short-lived low to high latitude
features.  The fragmented polar cap at epoch 1988.96 could indicate a
change in the nature of the dynamo in the star.  For the first time we
measure differential rotation for epochs with sufficient phase
coverage (1992.05, 1993.89, 1994.87).  These measurements show
variations on a timescale of at least one year, with the strongest
surface differential rotation ever measured for AB Dor occurring in
1994.86.  In conjunction with previous investigations, our results
represent the first long-term analysis of the temporal evolution of
differential rotation on active stars.

\end{abstract}


\section{Introduction}

Solar-type stars exhibit signatures of magnetic activity that are
assumed to be based on dynamo mechanisms operating in the star's outer
convective zone.  The detailed workings of the generation and
amplification mechanisms of the stellar dynamos are still poorly
understood as a result of the complex physics involved in theoretical
models, and the lack of observational constraints.  However, it is
widely accepted that differential rotation and convection are
essential ingredients of common theoretical amplification models
\citep{parker55,babcock65,leighton64,leighton69}.

Differential rotation results from the interplay of rotation and
convection, which leads to a redistribution of heat and angular
momentum inside the convection zone.  The thermal and density profiles
of convective motions produce a dependence of differential rotation on
both stellar latitude and radius.  It is a key process in the cyclic
activity of the stellar dynamo being the process through which
poloidal-toroidal field conversion occurs.  To date there have been
numerous measurements of differential rotation on rapidly rotating
cool stars, summarised by \cite{barnes05}, that show a steady increase
in the magnitude of differential rotation towards earlier spectral
types, consistent with the theoretical predictions of
\cite{ruediger02}.  In addition, helioseismology reveals a
differential rotation profile that varies with radius and latitude and
is caused by the presence of an intense turbulence in the solar
convention zone, which is driven by Reynolds stresses \citep{brun02}.

\begin{table*}
\begin{tabular}{l c c c c }
\hline
\hline

Epoch & Spectral Domain (nm) & Pixel Size (km\,s$^{-1}$) & Resolution & Telescope/Spectrograph \\
\hline

1988 Dec 16,19 & 605.9249 - 724.4182  & 2.7  & 56000 & AAT/UCL \\
1988 Dec 21  & 581.0809 - 673.5818  & 4.6 & 27000  & ESO(3.6m)/CASPEC\\
1992 Jan 18,19,20 & 498.2082 - 708.8381 & 2.42 & 51100 & AAT/UCL \\
1992 Dec 14 &  452.9687 - 688.6151 & 2.98 & 33600 &  AAT/UCL\\
1993 Nov 23,24,25 & 479.6900 - 751.7410 & 2.98 & 33600 & AAT/UCL \\
1994 Nov 15,16,17 & 551.0686 - 792.9818 &  3.56 & 26600 & CTIO \\

\hline
\hline

\end{tabular}
\caption{Journal of observations for AB Dor showing the observed spectral 
domain, pixel size, resolution and the instrument used for each epoch.
}
\label{tbl1}
\end{table*}

\begin{table*}
\begin{tabular}{l c c c c l }
\hline
\hline

Epoch & Julian Date & Phase Range & Signal-to-noise & No of lines & References \\

\hline

1988.96 & 7511.7607 - 7514.8768 & -0.37-(-0.174) : 0.381-0.672 & 70-100 & 80 & \cite{cameron90masses} \\ 
1988.97 & 7516.5617 - 7518.8339 & 0.961-1.517 & 280-300 & 290 & \cite{cameron90masses} \\
1992.05 & 8639.9234 - 8642.2515 & 0.127-0.782 : 0.038-0.722 : 0.98-0.628 & 50-110 & 945 & \cite{cameron94doppler} \\
1992.95 & 8970.9158 - 8971.2737 & 0.092-0.766 & 110-180 & 2134 & \cite{cameron95doppler} \\
1993.89 & 9314.9249 - 9317.2631 & -0.656--0.034 : -0.266-0.918 : 0.175-0.886 & 56-95 & 1515 & \cite{unruh95doppler} \\
1994.87 & 9671.5421 - 9673.8443 & 0.087-0.591 : 0.698-0.525 : 0.604-0.555 & 160-250 & 990 & \cite{cameron99} \\

\hline
\hline

\end{tabular}
\caption{Journal of observations showing Julian dates (+2450000), phase coverage, signal-to-noise of the data set, and the number of lines used in deconvolution.}
\label{tbl2}
\end{table*}


Several methods have been used to measure differential rotation.  The
methods of \cite{gray77,bruning81} and \cite{reiners02a} use Fourier
analysis to detect differential rotation through line profile
analysis.  These methods are only applicable to stars that do not
exhibit large cool starspots as they distort the shape of the line
profile.  Other methods reconstruct surface brightness images over a
time period and then trace surface features to ascertain how their
rotation periods are dependant on latitude.  Examples include, the
cross-correlation method used by \cite{donati97doppler} on AB Dor to
obtain the first differential rotation measurement for a star other
than the Sun.  This method measures the amount of rotational shear as
a function of latitude by cross-correlating belts of equal latitude.
The sheared-image method extends this to include a solar-like
differential rotation law into the image reconstruction process, where
the rotation rate is allowed to vary smoothly with latitude on an
image grid \citep{donati00rxj1508}.

The temporal evolution of differential rotation can be determined by
measurements over several epochs as shown for AB Dor by
\cite{cameron02twist}.  In this work, a matched-filter analysis method
was used to track individual spot features in the trailed
spectrograms.  The temporal evolution of differential rotation on AB
Dor has also been confirmed by \cite{donati03}, through the use of the
sheared-image method, though not for the same epochs as
\cite{cameron02twist}.



In this paper we extend the epochs for which differential rotation has
been measured using the sheared-image method, by processing archival
AB Dor data for epochs December 1988, January 1992, November 1993 and
November 1994 (presented in Section 2).  This is the first time that
surface brightness images have been reconstructed from composite
profiles computed from these data sets using the signal-to-noise
enhancement technique Least-Squares Deconvolution (LSD) (presented in
Sections 3 and 4).  Finally we measure differential rotation for each
eopch with sufficient rotational phase coverage (presented in Section
5) and discuss the implications of our results in Section 6.

\section{Observations and Data Modelling} 


The details of the instrument configuration and observing procedures
used to secure the six data sets are summarised in Table~\ref{tbl1}.
We refer the reader to the publications listed in Table~\ref{tbl2}
for further details.

\subsection{Data Reduction}

All frames were processed with ESpRIT, a dedicated package for the
optimal extraction of echelle spectroscopic observations.  Firstly a
3-D fit of the bias frame is subtracted from the flat-field and arc
frames.  Each order of the raw echelle frame is then located and
traced using cross-correlations with a user defined reference profile.
A linear or 2D fit to the shape of the arc lines provides the slit
direction.  Deviations from the slit direction are averaged over all
orders provide the mean slit shape.

The first step in the wavelength calibration procedure is to obtain an
accurate identification of calibration lines.  To start, the user
specifies the order numbers, and approximate values for the wavelength
of the first pixel and pixel size.  This information is used to
determine the location of lines from a calibration line list.  A
quadratic dispersion polynomial is then determined and used to
calibrate the remaining orders.  The final dispersion relation is
obtained by fitting a 2D polynomial to the pixel positions and
corresponding wavelengths of all lines simultaneously.  The comprising 
dimensions of the polynomial fit are; 1 dimension to fit the 
dispersion relation of each order and 1 dimension to fit its 
variation from one order to the next.

Pixel-to-pixel sensitivity differences are removed by dividing each
pixel in the stellar frame by the corresponding pixel in the
flat-field.  A 2D polynomial fit to the inter order background is
subtracted from the stellar frame.  All pixels that deviate from the
average intensity of the order (i.e. cosmic rays) are removed.  The
optimal extraction of \cite{marsh89} is then implemented.  The
continuum is automatically fitted firstly by using a high-degree 1D
polynomial, and then by a 2D polynomial to identify any systematic
trends in the continuum's shape.  A more detailed description of this
package is given by Donati et al. (1997).

\subsection {Least-Squares Deconvolution}

LSD is a method for combining the rotation profiles of thousands of
spectral lines in an optimally weighted manner (Donati et al. 1997).
It uses a weighted least squares algorithm to compute the line
broadening profile which, when convolved with the known pattern of
photospheric absorption lines in a stellar spectrum, optimises the
chi-squared fit to the data.  The list of spectral lines is obtained
from the LTE model atmospheres of \cite{kurucz93cdrom} for
T$_{eff}$=5000\,K and log\,g = 4.5, where features with a relative
central depth of at least 40\% of the local continuum flux are used.
The total number of lines used for each epoch is shown in
Table~\ref{tbl2}.  

It can be shown that the enhancement in signal to noise is equivalent
to either an optimally-weighted stacking of the profiles, or to
cross-correlation with the line pattern, scaling as the square root of
the number of lines used. It has the advantage, however, that
sidelobes caused by blends are automatically eliminated from the
composite profile, giving a clean profile surrounded by flat
continuum.  LSD conserves the shape of the rotational profile,
implying that any deviations in this profile can be interpreted as
brightness inhomogeneities on the stellar surface.

\begin{figure}
\psfig{file=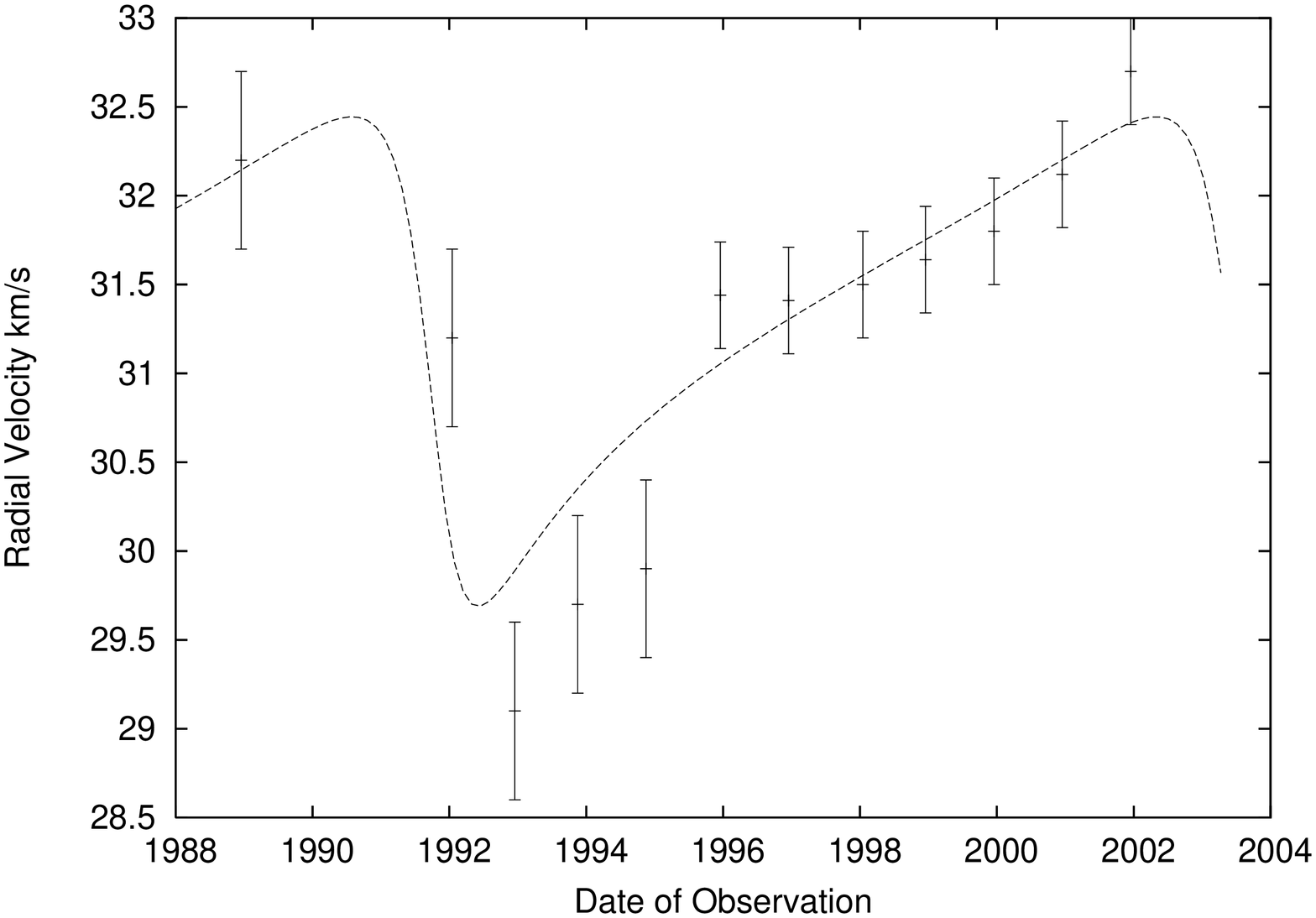,angle=270,width=8cm,height=6.5cm}
\caption{The orbital solution of \protect\cite{close05} for AB Dor A's
reflex orbit.  Also plotted are the empirically derived radial
velocity values for AB Dor as determined by this analysis (epochs 1988
to 1994), 
The orbital solution
includes an offset in the y-axis of 31.37\,km\,s$^{-1}$ that
corresponds to the radial velocity of the binary system, which was
determined by $\chi^2$ minimisation.  The $\chi^2$ obtained for this 
solution is 6.023.}
\label{f-radvel}
\end{figure}


\begin{figure*}
\hspace{-0.5cm}
\psfig{file=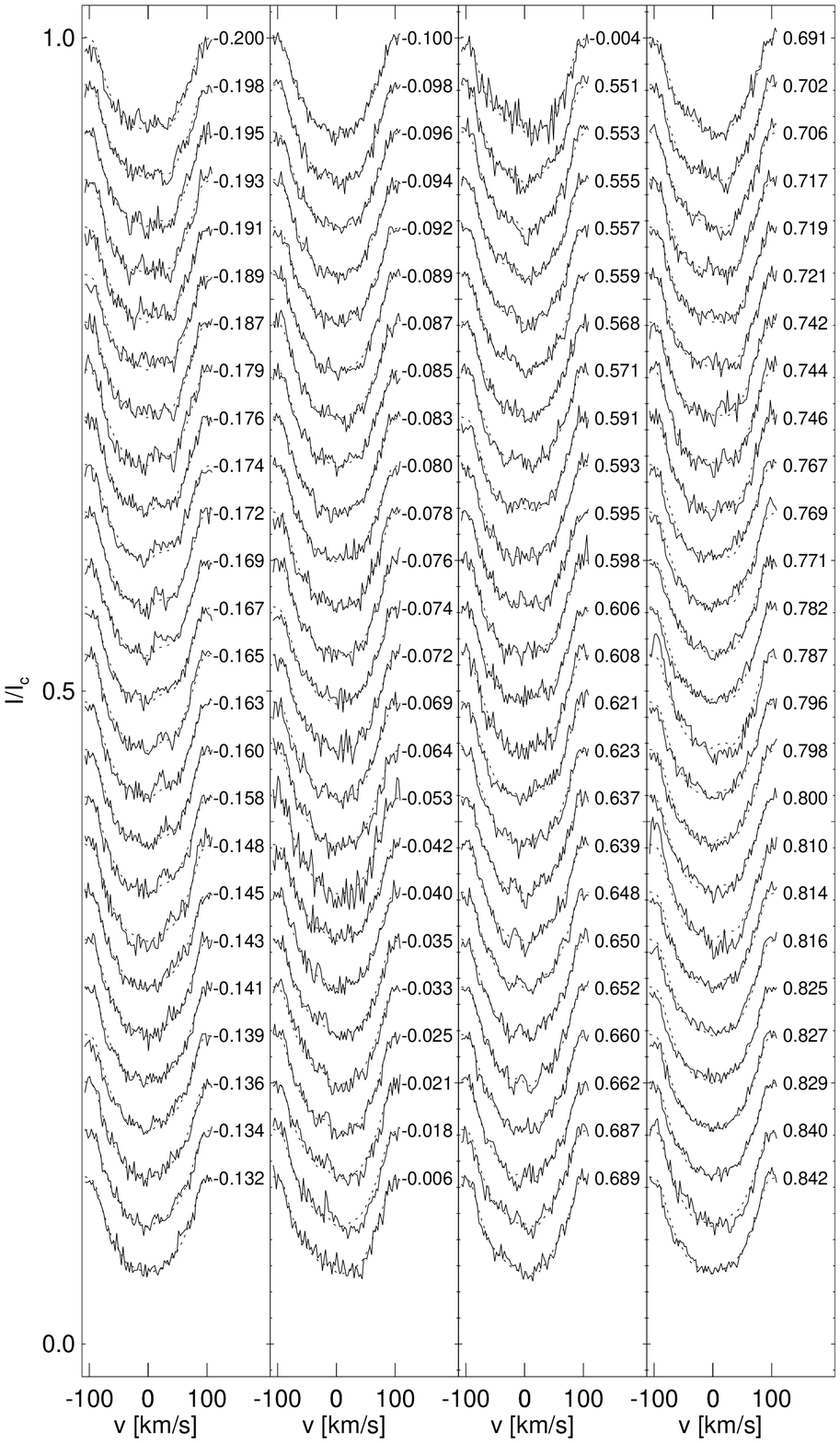,angle=0,width=15cm,height=20cm}
\caption{Maximum entropy fits (dashed line) to the LSD profiles (solid line) for 16+19 December 1988, observed at the AAT.  The rotational phases are indicated to the right of 
each profile.}
\label{fd88a}
\end{figure*}

\begin{figure*}
\hspace{-0.5cm}
\psfig{file=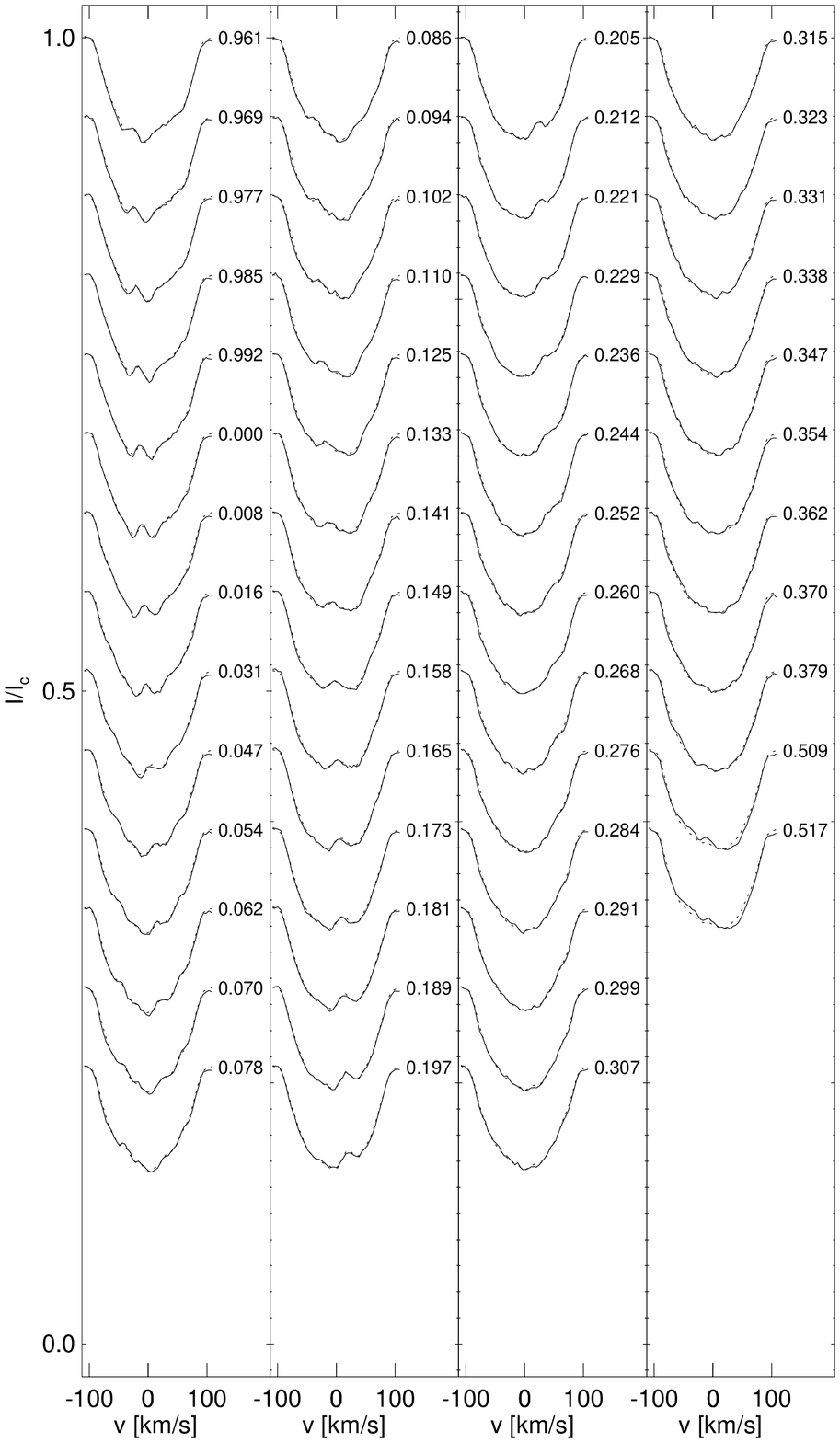,angle=0,width=12cm,height=14cm}
\caption{Maximum entropy fits (dashed line) to the LSD profiles (solid line) for 21 December 1988, observed at ESO.  The rotational phases are indicated to the right of 
each profile.}
\label{fd88b}
\end{figure*}


\begin{figure*}
\begin{tabular}{ll}

\psfig{figure=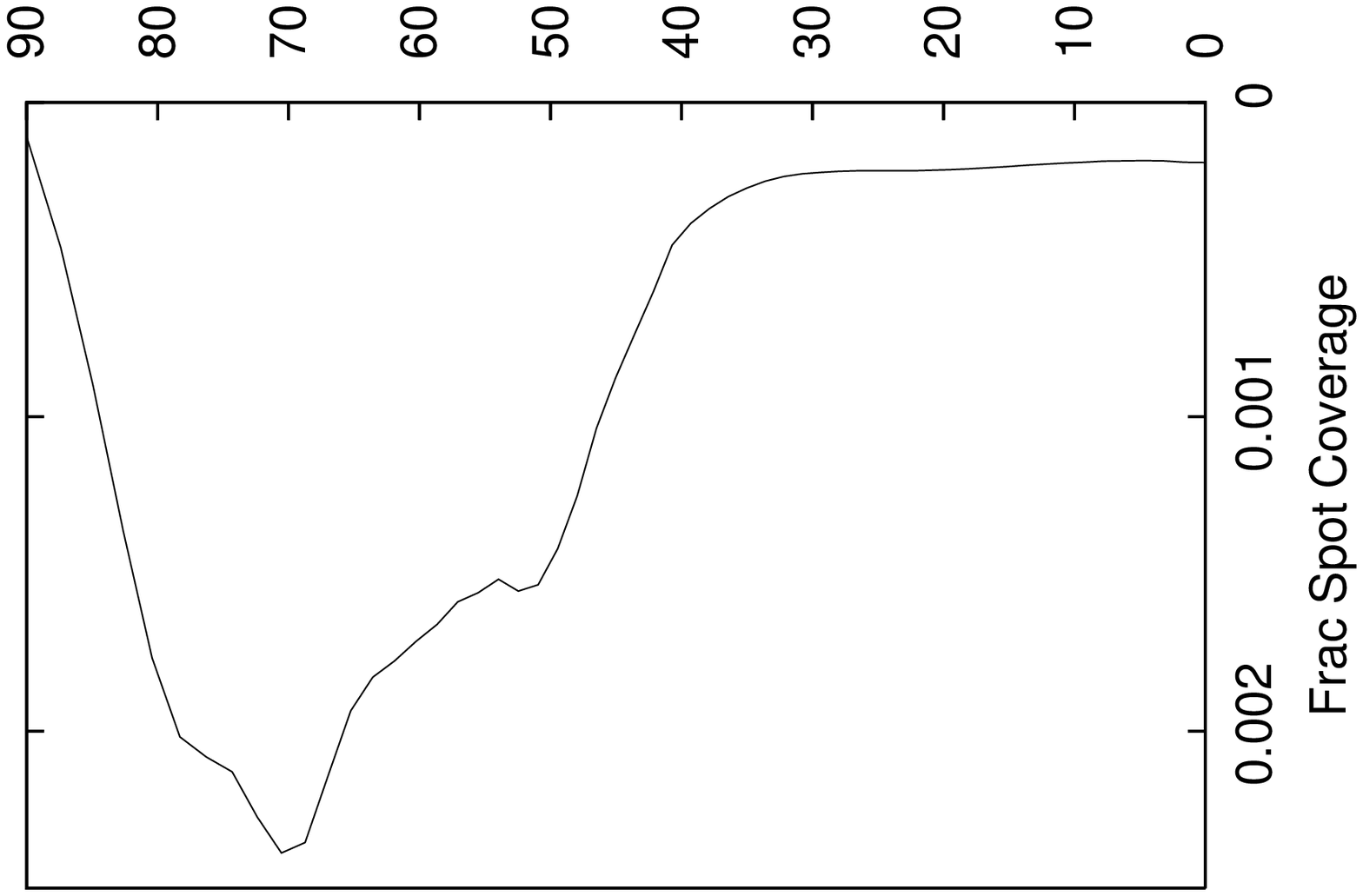,angle=270,width=2.5cm,height=5cm,bbllx=608bp,bblly=50bp,bburx=55bp,bbury=346bp} &
\psfig{figure=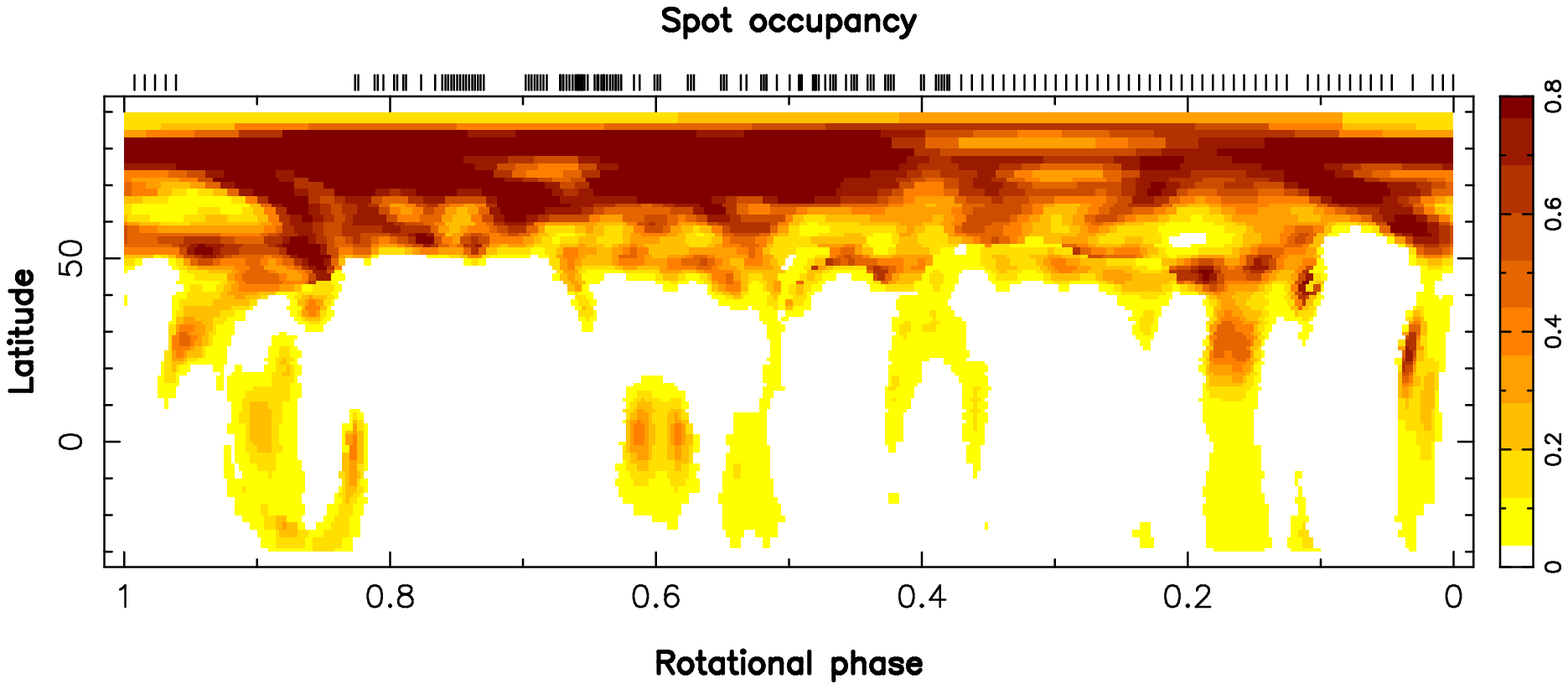,angle=0,width=13cm,height=6cm,bbllx=36bp,bblly=23bp,bburx=527bp,bbury=270bp} \\ \\
&
\hspace{-1.32cm}
\psfig{figure=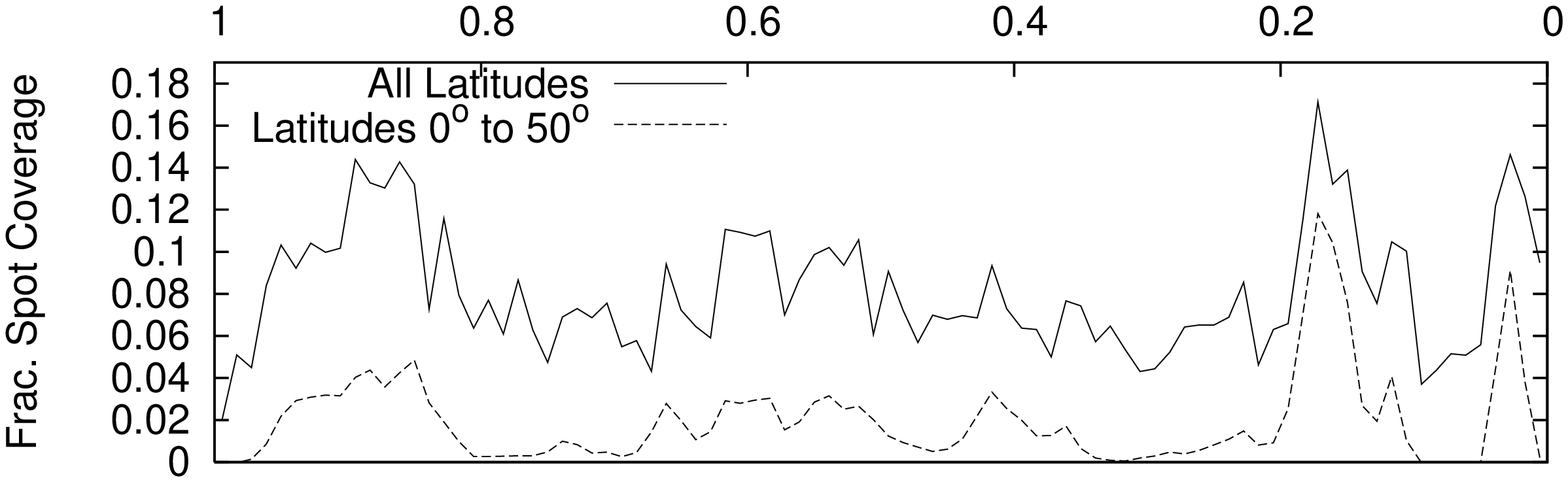,angle=270,width=15.4cm,height=3.5cm,bbllx=548bp,bblly=53bp,bburx=327bp,bbury=758bp} \\

\end{tabular}
\caption{Maximum Entropy surface brightness distribution for December 1988 
(epoch 1988.96), where the vertical tics at the top of the plot
indicate the phase coverage.  The plot to the left shows the
fractional spot coverage per latitude bin integrated over
longitude, while the plot below shows the fractional spot coverage per
rotational phase bin, integrated over latitude.}
\label{smd88}
\end{figure*} 



\begin{figure*}
\hspace{-0.5cm}
\psfig{file=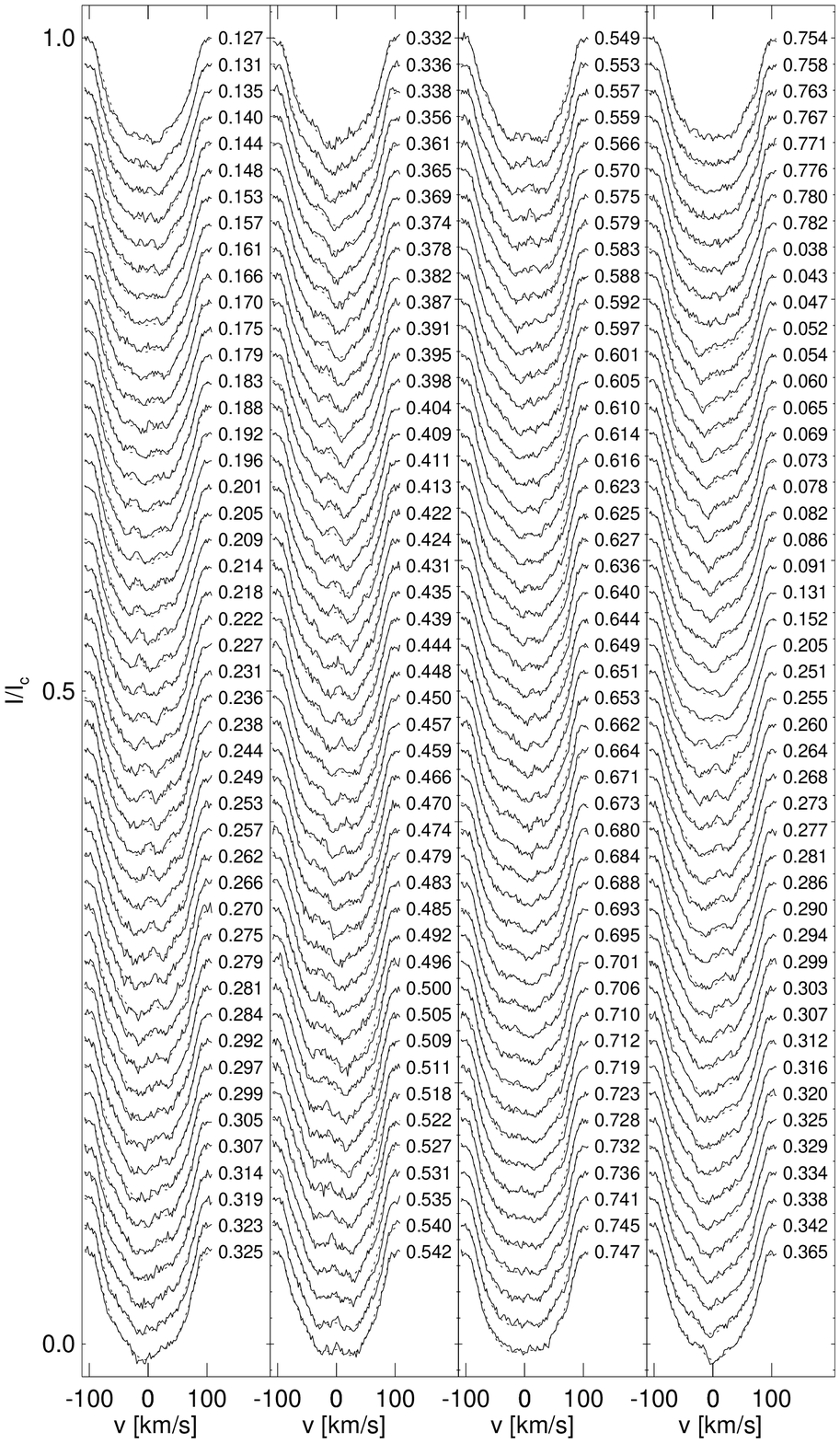,angle=0,width=15cm,height=20cm}
\caption{Maximum entropy fits (dashed line) to the LSD profiles (solid line) for 18-20 January 1992, part I.  The rotational phases are indicated to the right of each profile.}
\label{fj92a}
\end{figure*}

\begin{figure*}
\hspace{-0.5cm}
\psfig{file=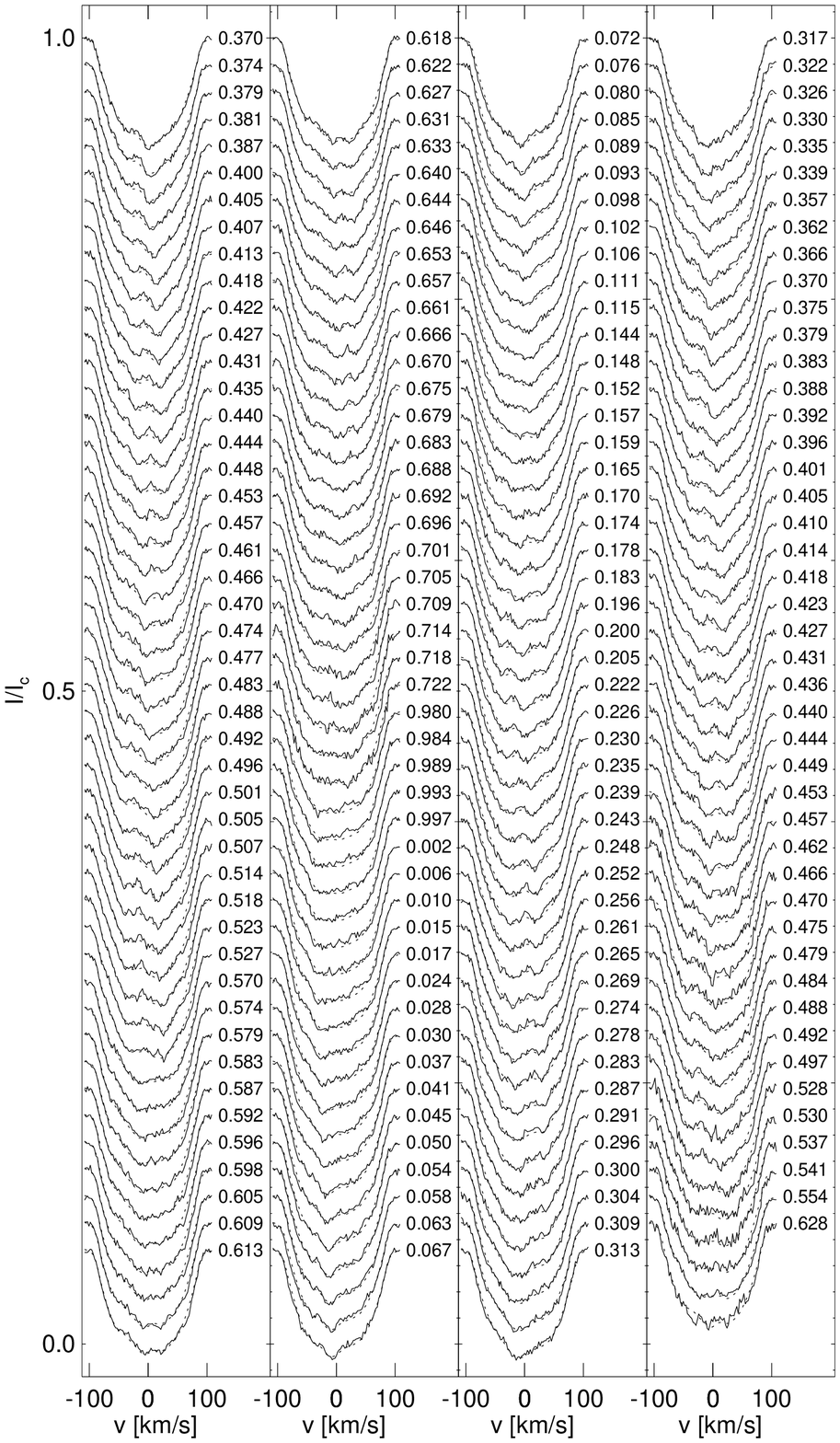,angle=0,width=15cm,height=20cm}
\caption{Maximum entropy fits (dashed line) to the LSD profiles (solid line) for 18-20 January 1992, part II.  The rotational phases are indicated to the right of each profile.}
\label{fj92b}
\end{figure*}


\begin{figure*}
\begin{tabular}{ll}

\psfig{figure=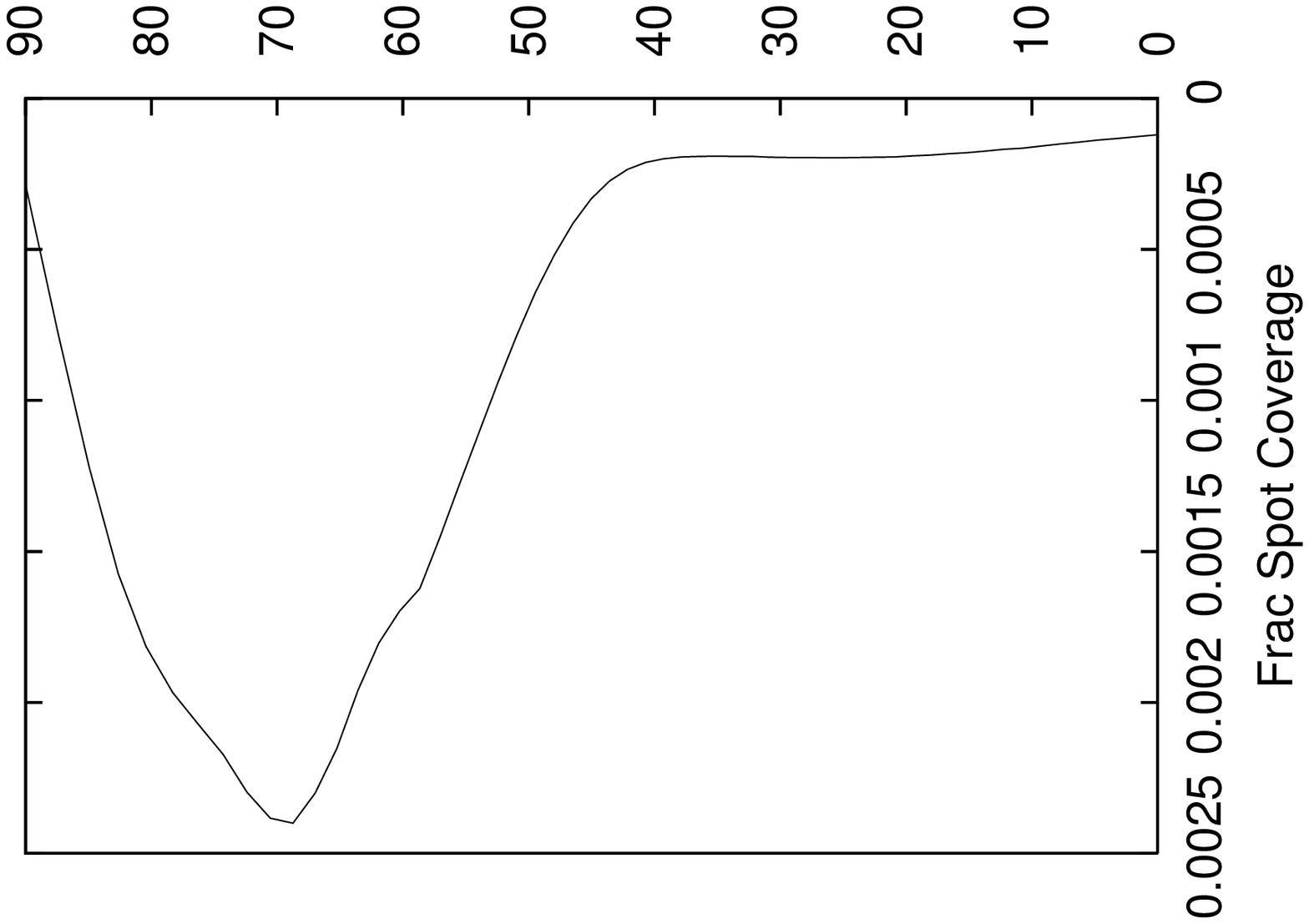,angle=270,width=2.5cm,height=5cm,bbllx=608bp,bblly=50bp,bburx=55bp,bbury=346bp} &
\psfig{figure=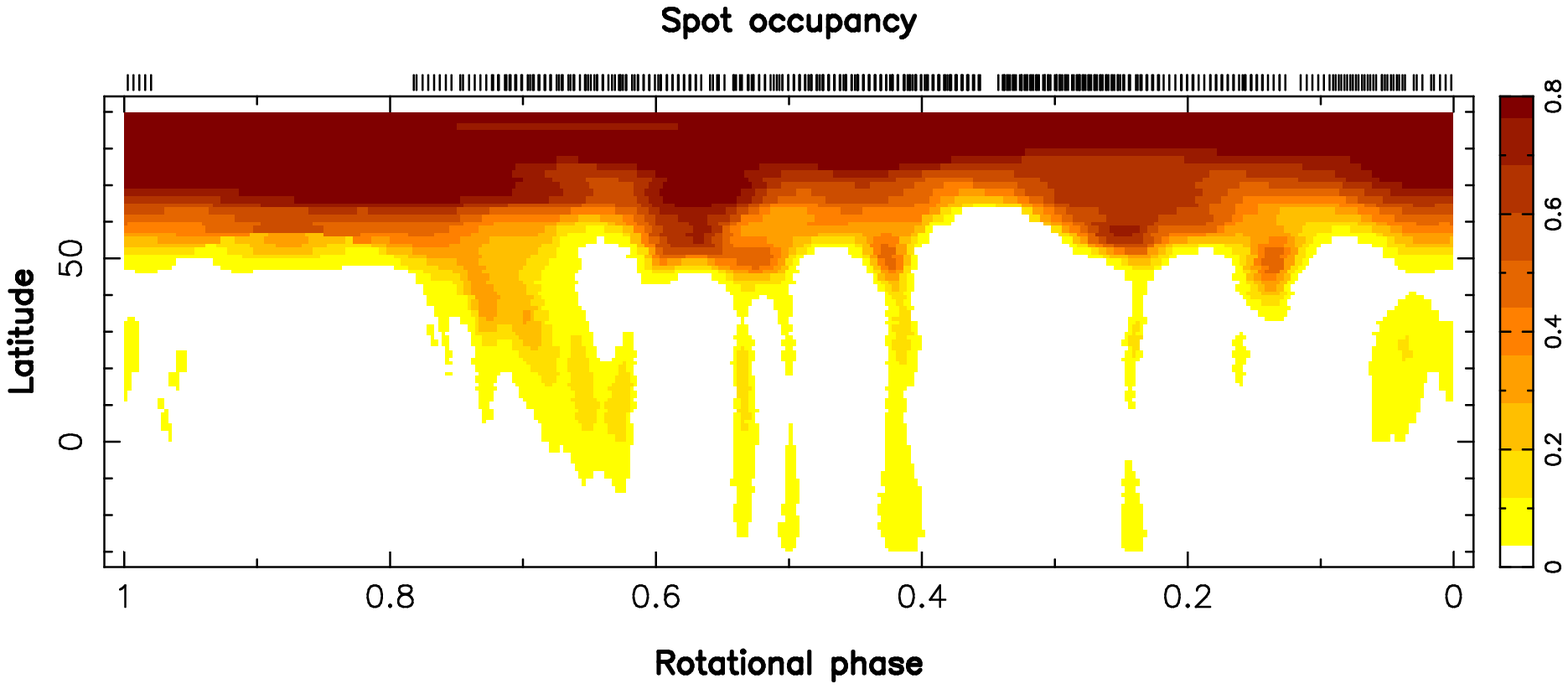,angle=0,width=13cm,height=6cm,bbllx=36bp,bblly=23bp,bburx=527bp,bbury=270bp} \\

&
\hspace{-1.32cm}
\psfig{figure=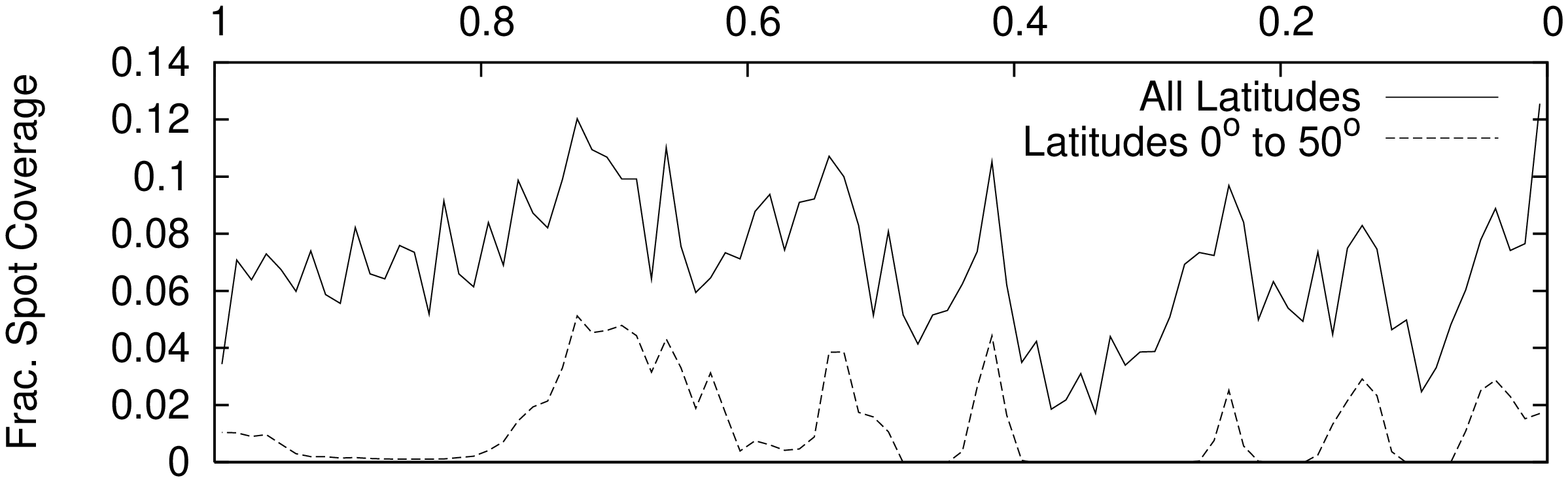,angle=270,width=15.4cm,height=3.5cm,bbllx=548bp,bblly=53bp,bburx=327bp,bbury=758bp} \\

\end{tabular}
\caption{Maximum Entropy surface brightness distribution for January 1992 
(epoch 1992.05), where the vertical tics at the top of the plot
indicate the phase coverage.  The plot to the left shows the
fractional spot coverage per latitude bin integrated over
longitude, while the plot below shows the fractional spot coverage per
rotational phase bin, integrated over latitude.}
\label{smj92}
\end{figure*} 


\begin{figure*}
\begin{tabular}{ll}

\psfig{figure=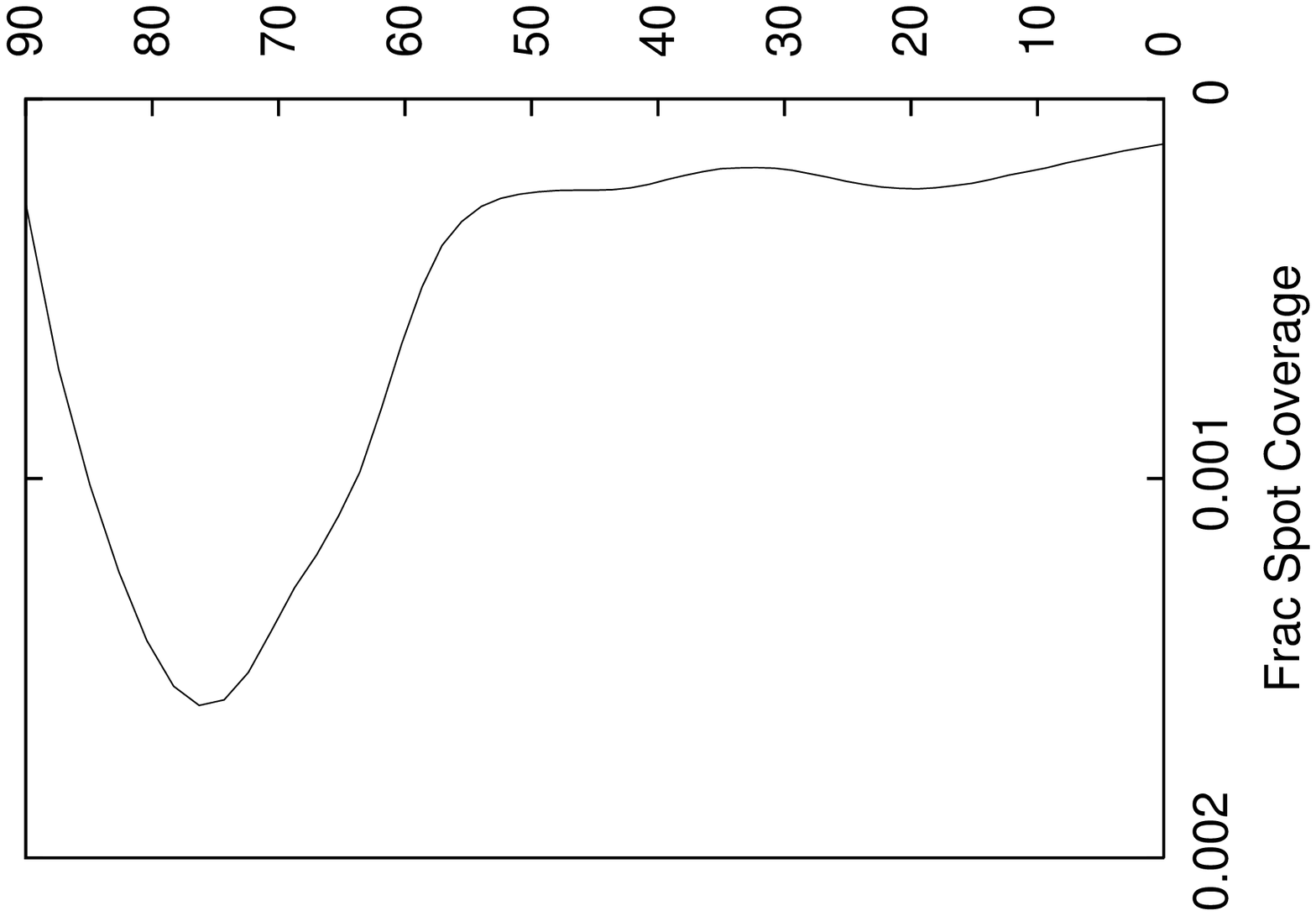,angle=270,width=2.5cm,height=5cm,bbllx=608bp,bblly=50bp,bburx=55bp,bbury=346bp} &
\psfig{figure=out_dec92_std_ph2.ps,angle=0,width=13cm,height=6cm,bbllx=36bp,bblly=23bp,bburx=527bp,bbury=270bp} \\

&  
\hspace{-1.32cm}
\psfig{figure=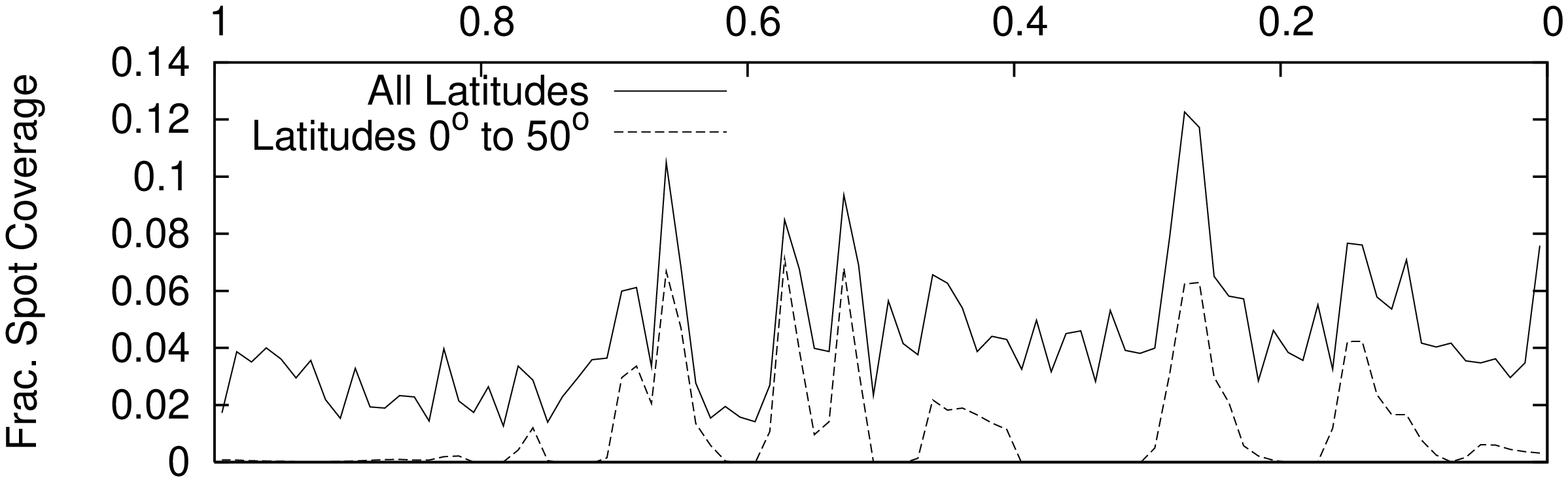,angle=270,width=15.4cm,height=3.5cm,bbllx=548bp,bblly=53bp,bburx=327bp,bbury=758bp} \\

\end{tabular}
\caption{Maximum Entropy surface brightness distribution for December 1992 
(epoch 1992.95), where the vertical tics at the top of the plot
indicate the phase coverage.  The plot to the left shows the
fractional spot coverage per latitude bin integrated over
longitude, while the plot below shows the fractional spot coverage per
rotational phase bin, integrated over latitude.}
\label{smd92}
\end{figure*}

\begin{figure*}
\hspace{-0.5cm}
\psfig{file=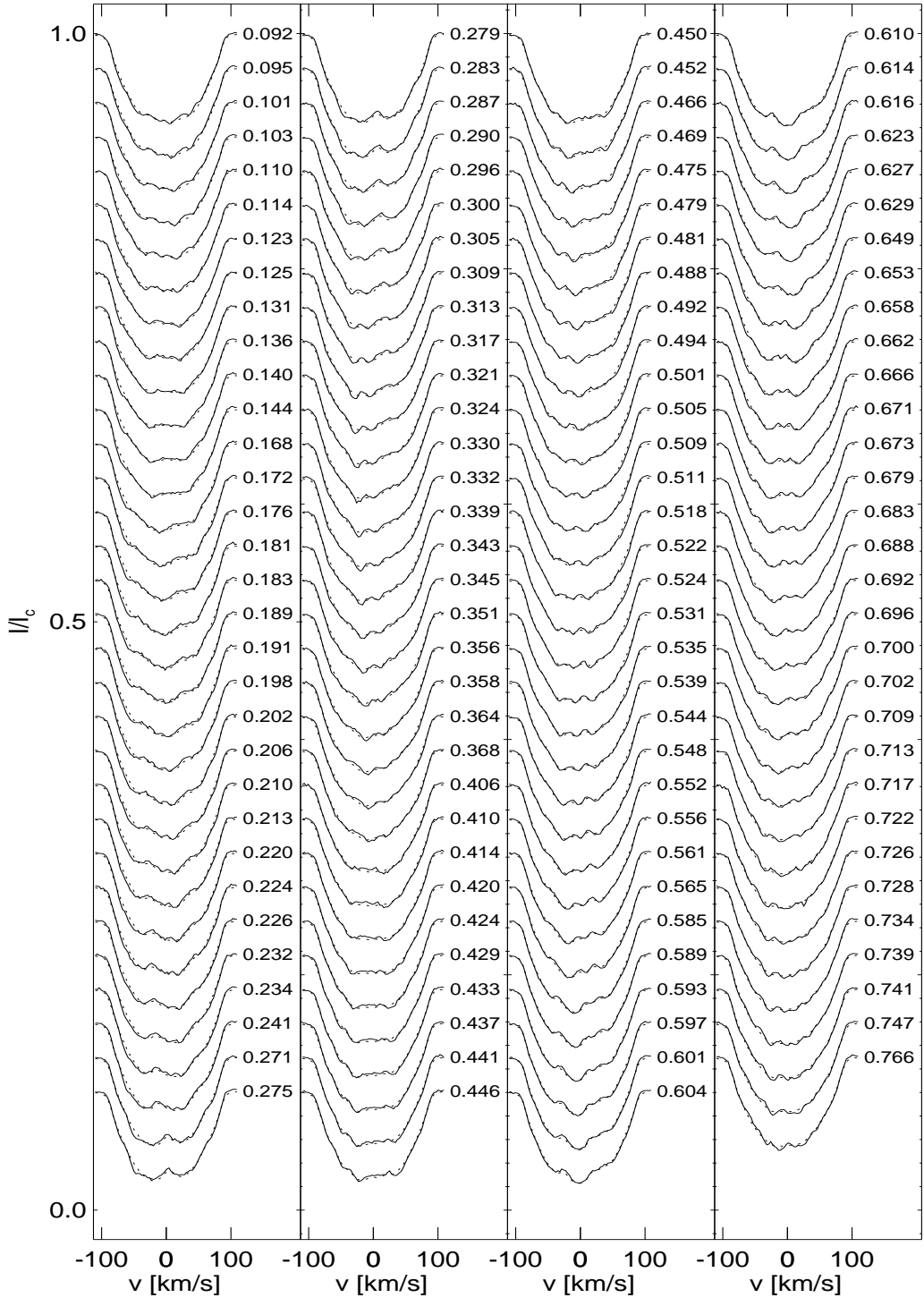,angle=0,width=15cm,height=20cm}
\caption{Maximum entropy fits (dashed line) to the LSD profiles (solid line) for 14 December 1992.  The rotational phases are indicated to the right of each profile.}
\label{fd92a}
\end{figure*}



\begin{figure*}
\hspace{-0.5cm}
\psfig{file=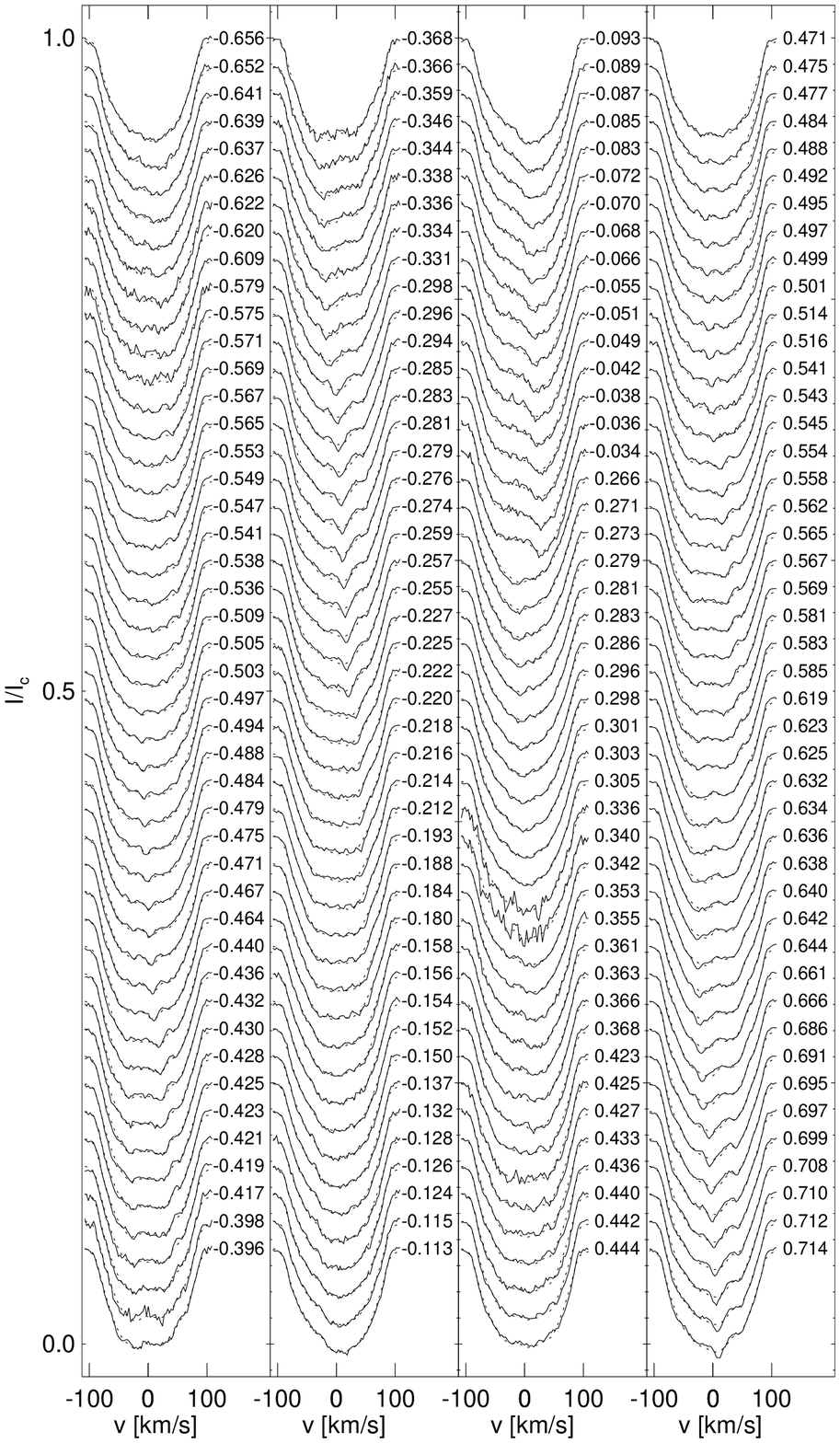,angle=0,width=15cm,height=20cm}
\caption{Maximum entropy fits (dashed line) to the LSD profiles (solid line) for 23-25 November 1993, part I.  The rotational phases are indicated to the right of each profile.}
\label{fn93a}
\end{figure*}

\begin{figure*}
\hspace{-0.5cm}
\psfig{file=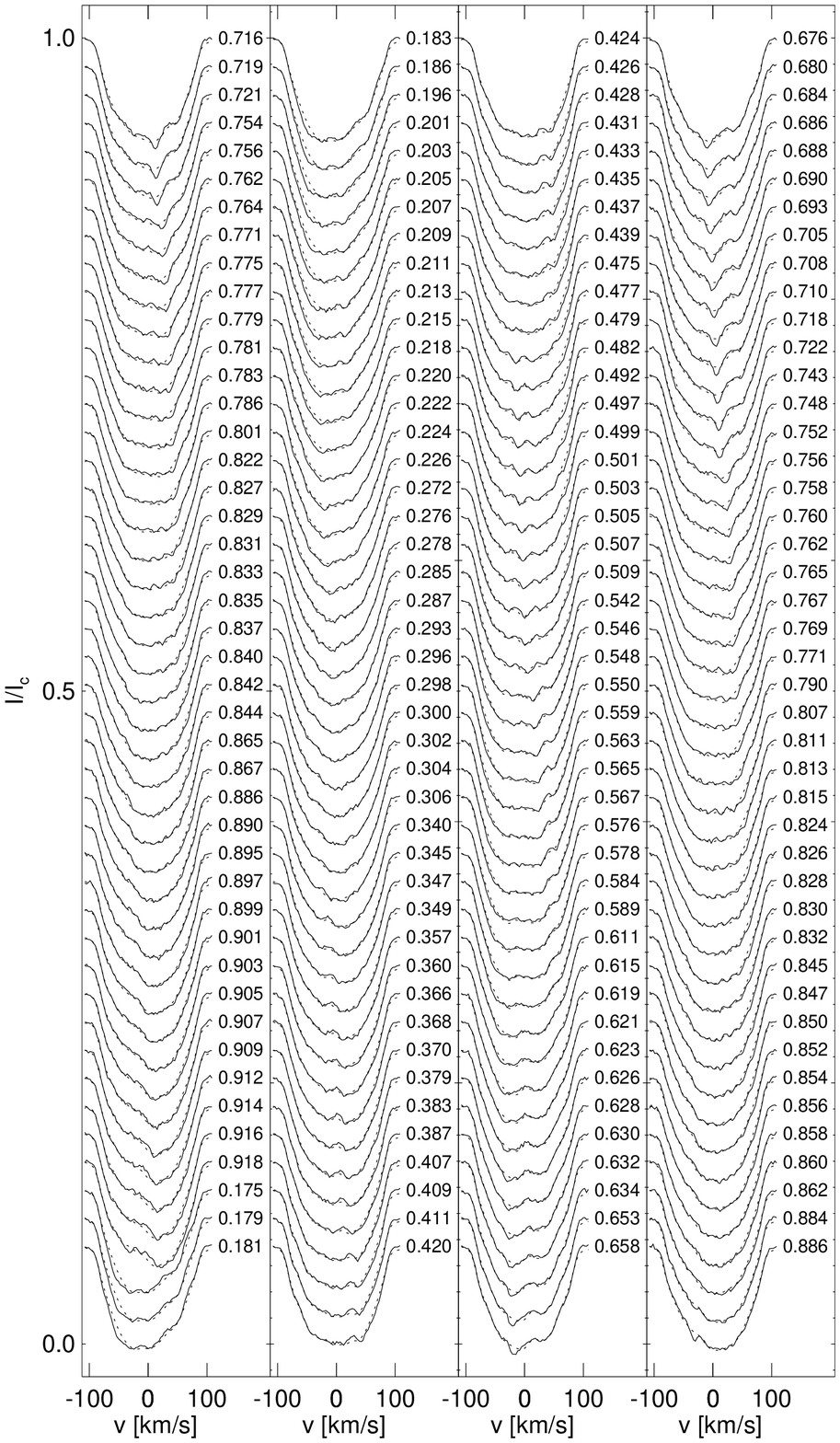,angle=0,width=15cm,height=20cm}
\caption{Maximum entropy fits (dashed line) to the LSD profiles (solid line) for 23-25 November 1993, part II.  The rotational phases are indicated to the right of each profile.}
\label{fn93b}
\end{figure*}


\begin{figure*}
\begin{tabular}{ll}

\psfig{figure=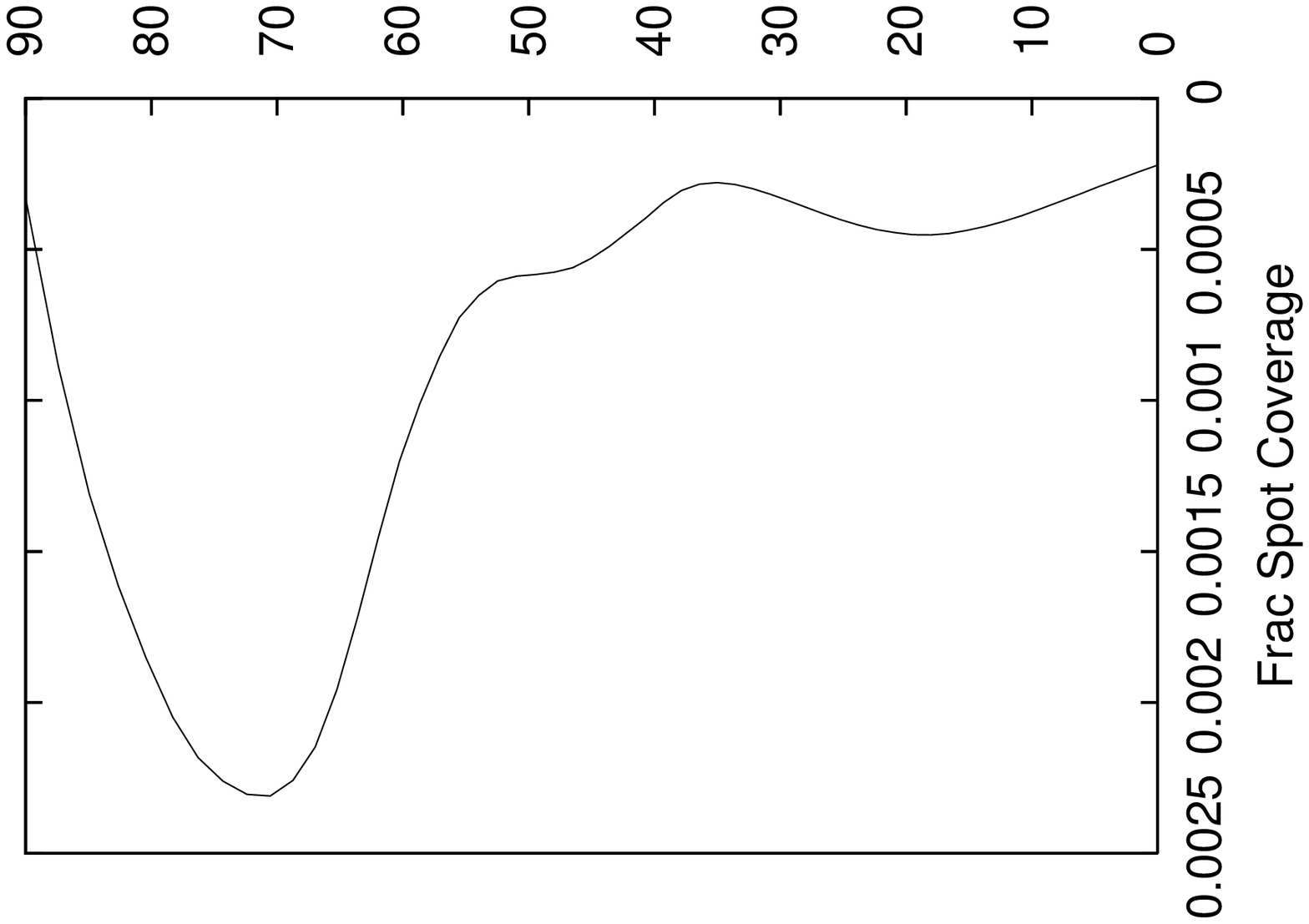,angle=270,width=2.5cm,height=5cm,bbllx=608bp,bblly=50bp,bburx=55bp,bbury=346bp} &
\psfig{figure=output_nov93_ph.ps,angle=0,width=13cm,height=6cm,bbllx=36bp,bblly=23bp,bburx=527bp,bbury=270bp} \\

&  
\hspace{-1.32cm}
\psfig{figure=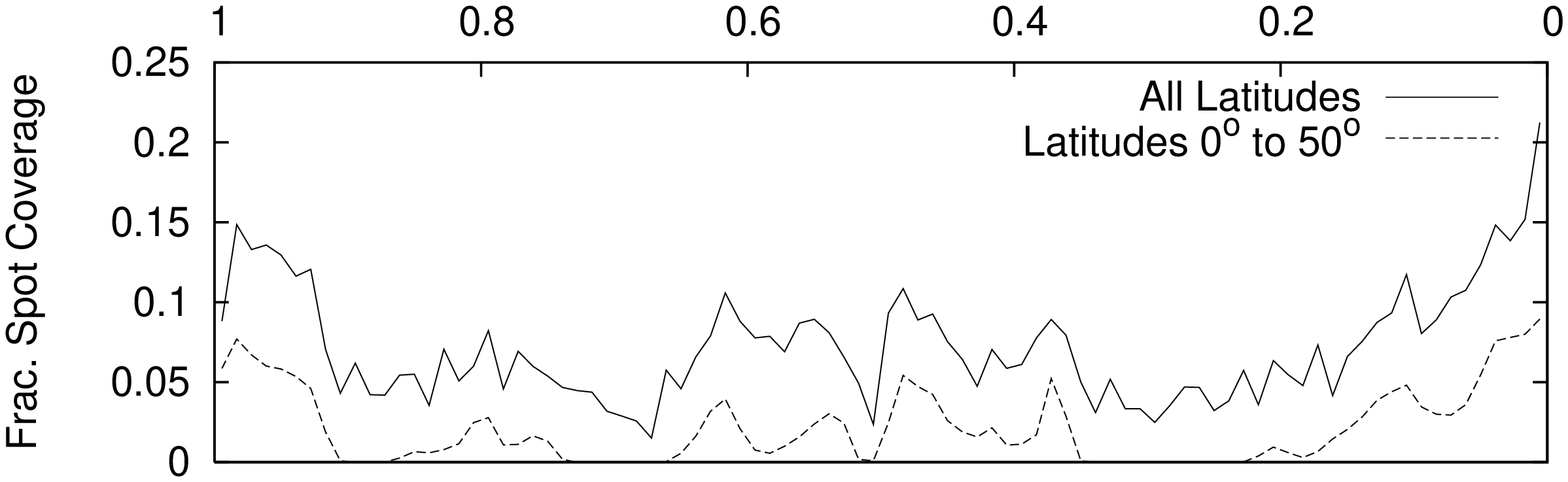,angle=270,width=15.4cm,height=3.5cm,bbllx=548bp,bblly=53bp,bburx=327bp,bbury=758bp} \\

 \end{tabular}
\caption{Maximum Entropy surface brightness distribution for November 1993 
(epoch 1993.89), where the vertical tics at the top of the plot
indicate the phase coverage.  The plot to the left shows the
fractional spot coverage per latitude bin integrated over
longitude, while the plot below shows the fractional spot coverage per
rotational phase bin, integrated over latitude.}
\label{smn93}
\label{image}
\end{figure*} 


\begin{figure*}
\hspace{-0.5cm}
\psfig{file=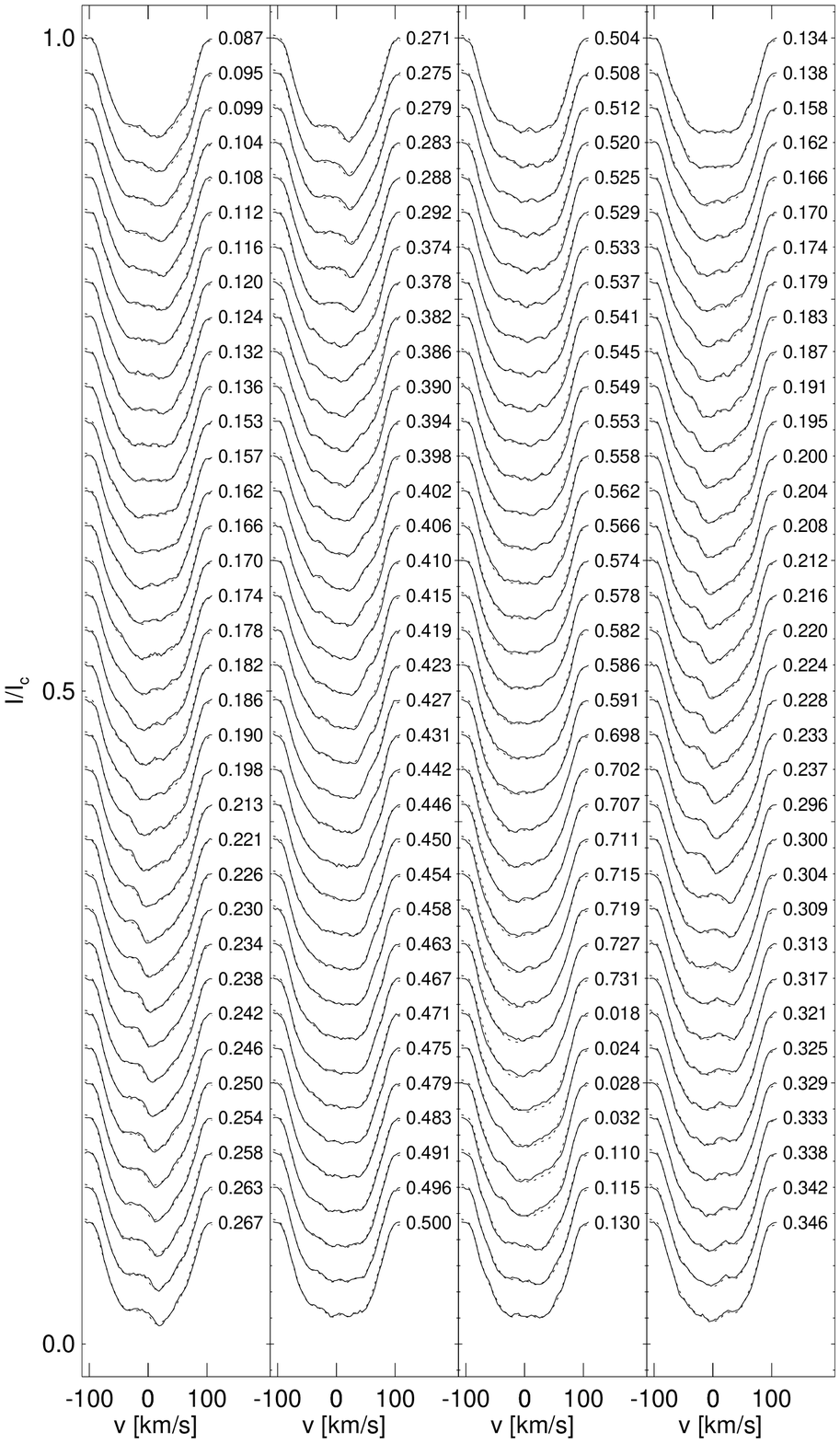,angle=0,width=15cm,height=20cm}
\caption{Maximum entropy fits (dashed line) to the LSD profiles (solid line) for 15-17 November 1994, part I.  The rotational phases are indicated to the right of each profile.}
\label{fn94a}
\end{figure*}

\begin{figure*}
\hspace{-0.5cm}
\psfig{file=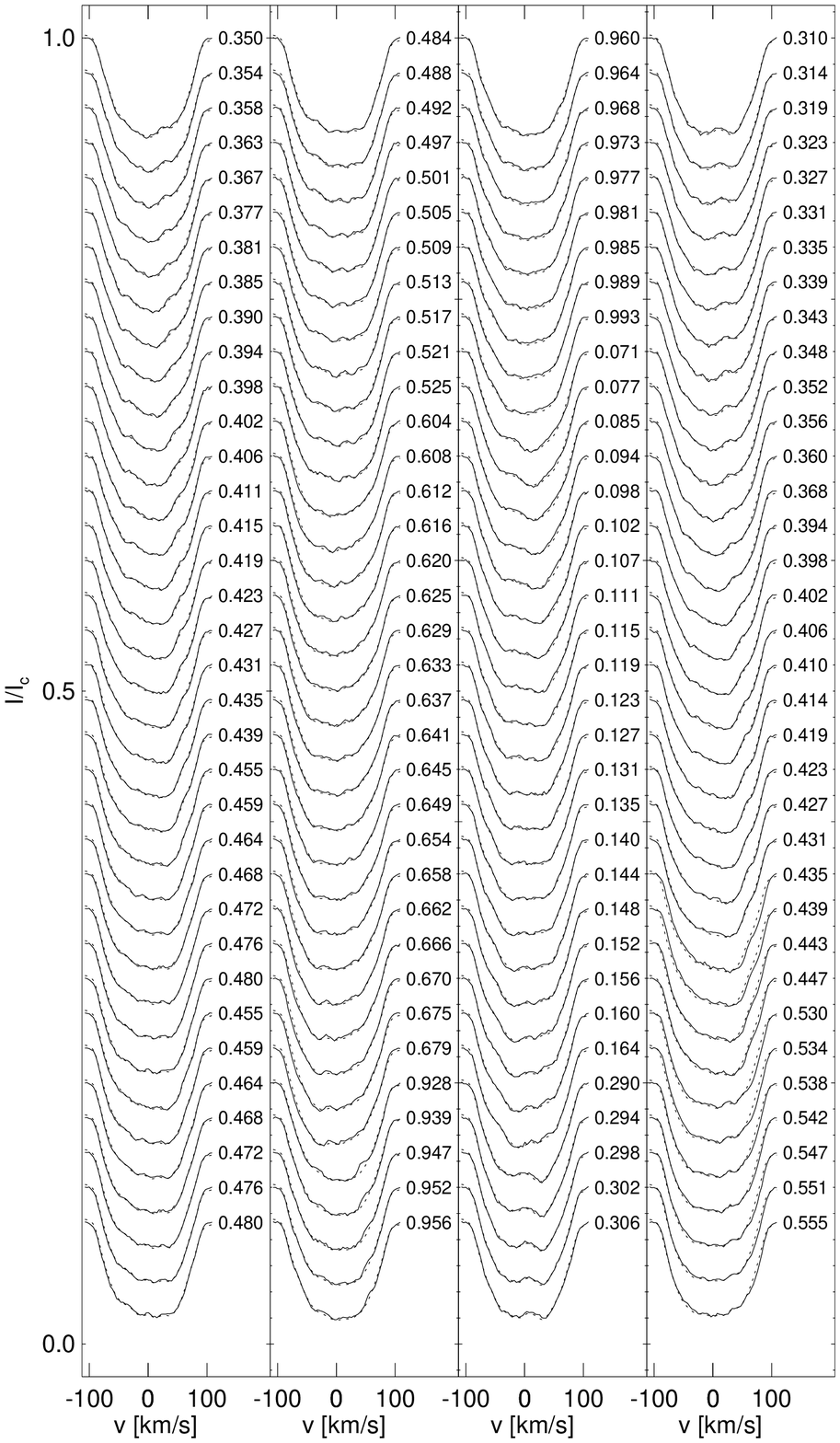,angle=0,width=15cm,height=20cm}
\caption{Maximum entropy fits (dashed line) to the LSD profiles (solid line) for 15-17 November 1994, part II.  The rotational phases are indicated to the right of each profile.}
\label{fn94b}
\end{figure*}


\begin{figure*}
\begin{tabular}{ll}

\psfig{figure=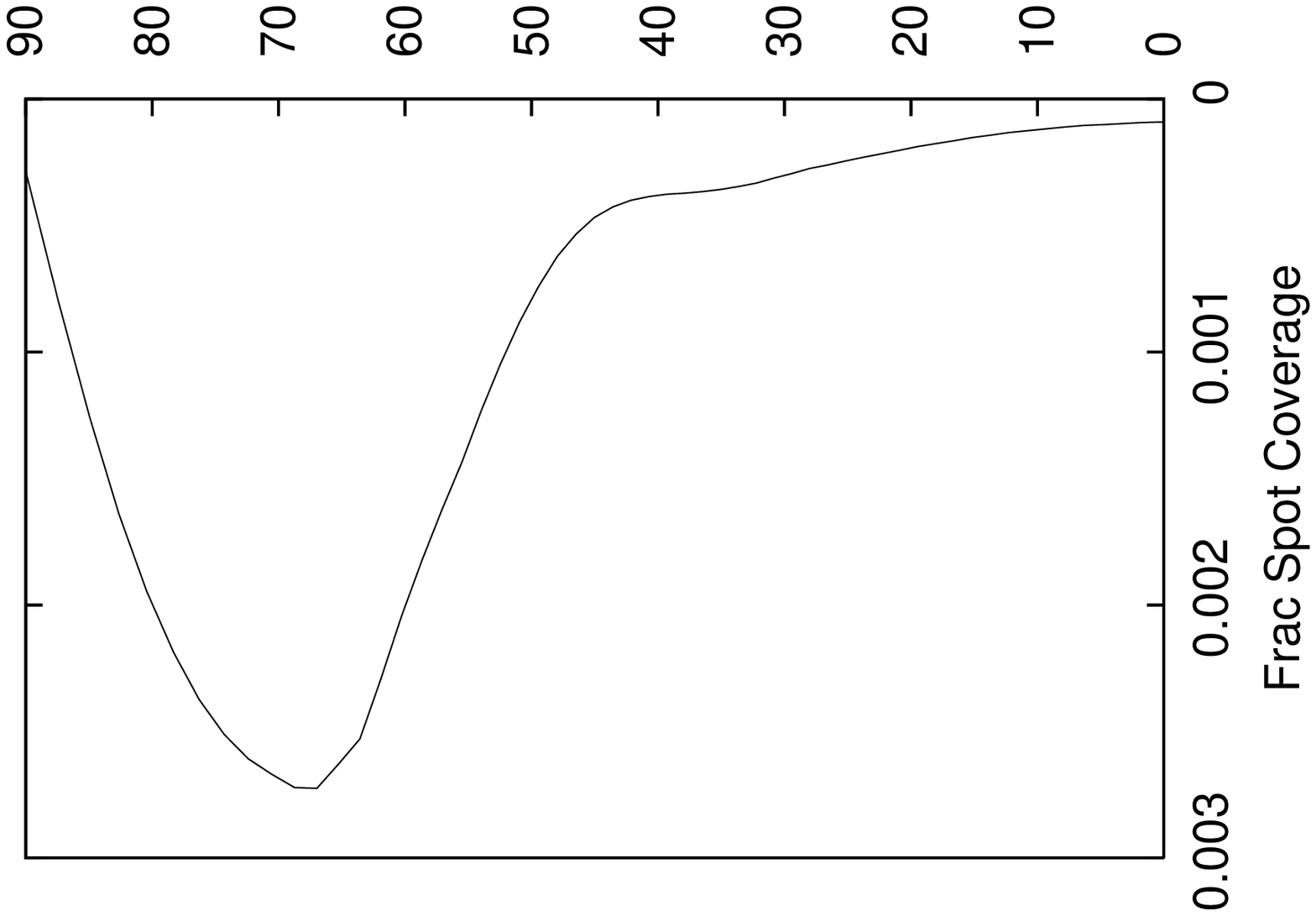,angle=270,width=2.5cm,height=5cm,bbllx=608bp,bblly=50bp,bburx=55bp,bbury=346bp} &
\psfig{figure=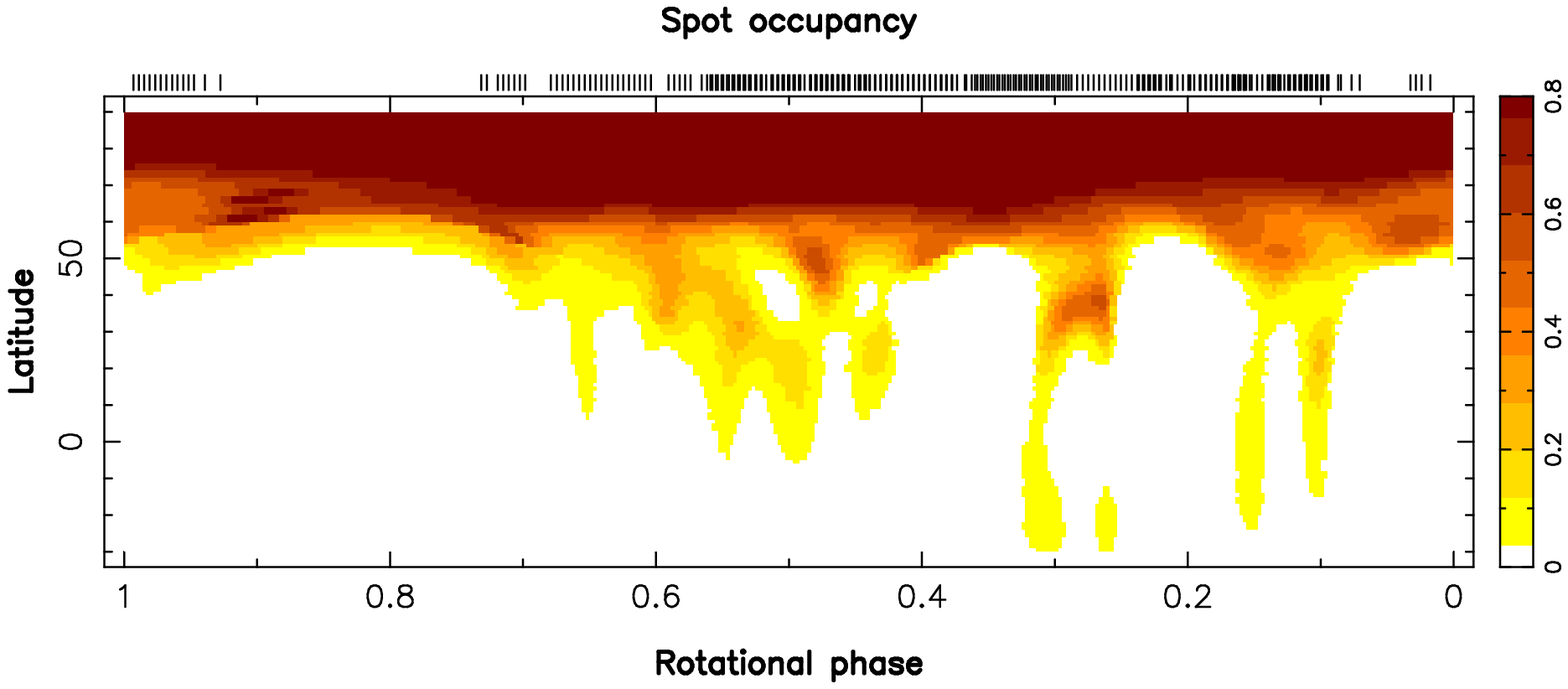,angle=0,width=13cm,height=6cm,bbllx=36bp,bblly=23bp,bburx=527bp,bbury=270bp} \\

&  
\hspace{-1.32cm}
\psfig{figure=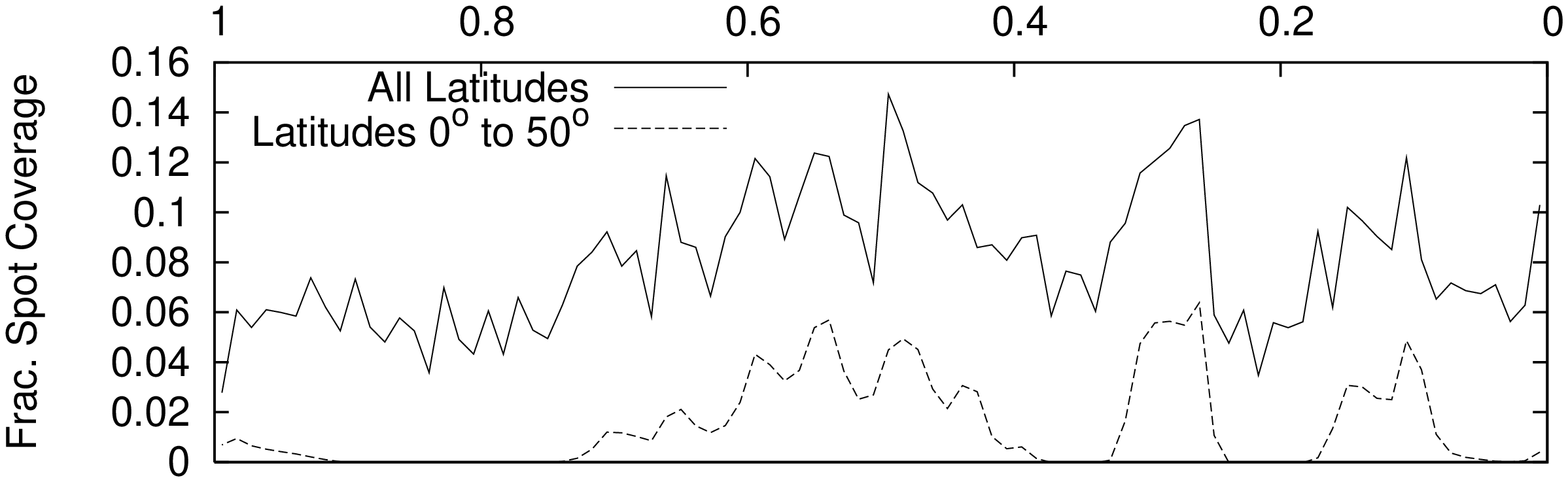,angle=270,width=15.4cm,height=3.5cm,bbllx=548bp,bblly=53bp,bburx=327bp,bbury=758bp} \\

 \end{tabular}
\caption{Maximum Entropy surface brightness distribution for November 1994
(epoch 1994.87), where the vertical tics at the top of the plot
indicate the phase coverage.  The plot to the left shows the
fractional spot coverage per latitude bin with integrated over
longitude, while the plot below shows the fractional spot coverage per
rotational phase bin, integrated over latitude.}
\label{smn94}
\label{image}
\end{figure*}

\section{Radial Velocity Correction}

\subsection{Telluric line alignment}

Telluric lines are used to correct for small shifts in the
spectrograph during the night i.e. from dewar refill and the thermal
and mechanical relaxation of spectrograph's components, as they are
only present in the Earth's atmosphere and are therefore at zero
radial velocity.  We use the procedure of \cite{donati03} in this
analysis.  Firstly the composite profile of telluric absorption lines
in each stellar spectrum is computed using LSD, with a line mask
comprising the wavelengths and relative strengths of known telluric
features (mainly of water) in the observed spectral region.  The
instrumental velocity of the composite telluric profile is then
measured and used to define the zero-point of the velocity scale.
Each stellar profile is then shifted by the measured instrumental
velocity before applying heliocentric velocity corrections.

\subsection{Radial Velocity Results}

The radial velocity derived for each epoch is tabulated in
Table~\ref{f-radvel}, where over the time-span of our observations
there is a variation of 3.4 km\,s$^{-1}$.  The change in radial
velocity reflects the orbital motion of AB Dor predominantly due to
the presence of its closest companion, AB Dor C (\cite{close05}).  AB
Dor C is a low mass object (0.09\,M$_\odot$) first detected by
\cite{guirado97}, which orbits AB Dor A at a separation of 0.156
arcseconds.  In addition to AB Dor C, AB Dor is known to have a wide
companion AB Dor B \citep{vilhu91,martin95} which is itself a close
binary system, and is separated from AB Dor A by a distance of
9.09$\pm$0.01 arcseconds.

The AB Dor A and C orbital solution has been determined by
\cite{close05}, where the parameters that define AB Dor A's reflex
orbit are fitted by: period\,=\,11.75\,$\pm$\,0.25\,yr, semi-major
axis\,=\,0.032\,$\pm 0.002''$, eccentricity\,=\,0.59\,$\pm$0.03,
periastron passage at 1991.8\,$\pm$\,0.2, inclination\,=\,67\,$\pm 3
^\circ$, $\Omega$=132\,$\pm 2 ^\circ$, $\omega$=107\,$\pm 7 ^\circ$.
The orbital solution is shown in Fig.~\ref{f-radvel} to fit the data
points within the error bars and to follow the decrease in radial
velocity at periastron (epoch 1991.8).  The error is approximately
0.5\,km\,s$^{-1}$ for the epochs of this work and 0.3\,km\,s$^{-1}$
for epochs after 1995, and results from broad and time variable
stellar lines.  The exceptions are the radial velocity measurements of
epochs 1992.95, 1993.89, and 1994.87, but the good fits to the line
profile shown in Figures~\ref{fj92a} \&~\ref{fj92b} for epoch 1992.05,
Figure~\ref{fd92a} for epoch 1992.95 and Figures~\ref{fn93a}
\&~\ref{fn93b} for epoch 1993.89, exclude any additional measurement
errors.  More empirical data points will help to constrain the errors
of the orbital solution.

\begin{table}
\hspace{0.5cm}
\begin{tabular}{l c}
\hline
\hline

Epoch & Radial Velocity (km\,s$^{-1}$) \\
\hline

1988 Dec 16 \& 19 & 32.2 \\
1992 Jan  & 31.2 \\
1992 Dec  & 28.8 \\
1993 Nov  & 29.7 \\
1994 Nov  & 29.9 \\
1995 Dec  & 31.4 \\
1996 Dec  & 31.4 \\
1998 Jan  & 31.5 \\
1998 Dec  & 31.6 \\
1999 Dec  & 31.8 \\
2000 Dec  & 32.1 \\
2001 Dec  & 32.7 \\
2002 Dec  & 32.9 \\

\hline
\hline

\end{tabular}
\caption{Radial velocity measurements for each epoch of this analysis
with previously published values for epochs after 1994.}
\label{t-radvel}
\end{table}

\section{Surface Image Reconstruction}

The surface brightness images are reconstructed using the maximum
entropy code of \cite{brown91zdi} and \cite{donati97recon}.  The
brightness model that is incorporated in this code is the `spot
occupancy' model of Collier Cameron (1992), where each point on the
stellar surface is quantified by the local fraction of the stellar
surface occupied by spots.  The range of spot occupancy is from 0,
where there are no spots present, to 1, where there is maximum
spottedness.

The imaging parameters that are used in this work for AB Dor are
stellar axial inclination, {\em i}\,=\,60$^o$, the projected
equatorial rotation velocity, $v$\,sin\,$i$ = 89\,km\,s$^{-1}$ and the
photospheric and spot temperatures respectively 5000\,K and 3500\,K
\citep{donati03}.  All data sets are phased according to the ephemeris 
of \cite{innis88II}; HJD = 244\,4296.575 + 0.51479\,E.  LSD profiles of
slowly rotating standard stars (GL 176.3 and GL 367) are used as template
profiles that represent the contribution of the photosphere and the spot
to the shape of the intrinsic line profile.

\subsection{Results}

All data sets provide good sampling of the rotational cycle of AB
Dor.  The maximum entropy images are structurally very similar with a
polar cap and/or high latitude spots, and with varying degrees of low
latitude spots.  The longitude resolution is approximately 3$^\circ$ at the 
equator and the size of the smallest features in latitude.

The first surface brightness image reconstructed for December 1988 is
shown in Figure~\ref{smd88}.  It shows high latitude spots that
dominate over a weak polar cap and a dearth of low latitude spot
coverage.  These sets were originally observed with the aim of
detecting circumstellar clouds on AB Dor using H$\alpha$, Ca {\sc
ii}\,H and K, and Mg {\sc ii} h and k lines
\cite{cameron90masses}.  The most striking feature of this data set is
the fragmented polar cap.  However, this could result from the low S/N
of the data set.

The reconstructed surface-brightness image for January 1992 is shown
in Figure~\ref{smj92}. 
The image is comparable with previous images of this data set
processed without LSD \cite{cameron94doppler} using the Ca I 643.9nm
and Ca 671.8 nm photospheric lines.  It is not possible to compare
exact features due to the signal-to-noise difference, but to a first
approximation the two images contain similar spot features at mid to
low latitudes.  Examples of such features are at phase 0.65, the gap
in low to mid-latitude features at phase 0.35, and the spot features
at phase 0.5.  \cite{cameron94doppler} also reconstructed a surface
brightness image using the photospheric line Fe I 666.3nm.  However,
the reconstructed spot features and groupings are less similar to the
image shown in Figure~\ref{smj92}, which is likely to result from the
different excitations of the two lines.  At high to polar latitudes the
images reconstructed in this analysis show a strong and uniform polar
cap whereas the images of
\cite{cameron94doppler} show weak and fragmented spot structures.  It
should be noted that the default value of spot coverage is 0.5 in the
work of \cite{cameron94doppler}, while in this analysis we set the
value to be 0.999.  A surface image has also been reconstructed by 
\cite{jarvinen05} using photometric data for the epoch 1991.96, which 
shows a primary spot at phase 0.7 and a weaker secondary spot at phase
1.097.  While Figure~\ref{smj92} shows that there is a fragmented
spot structure at phases 0.6 to 0.8, there is no evidence for a second
grouping of spots at phase 1.097.

For the single night of observations in December 1992 
the reconstructed surface brightness image is broadly in agreement
with that previously reconstructed by
\cite{cameron95doppler} using Ca I 643.9\,nm, Fe I 666.3\,nm and
Ca I 671.8\,nm.  Both images show an off-centre polar cap though the
spot structure is less fragmented, and the lower latitude features are
more clearly resolved in the surface brightness images reconstructed
in this analysis.  Examples of common features include spots at phases
0.15, 0.28 and 0.45, and a large unspotted area at phase 0.35.  For
epoch 1992.96 the photometric image reconstruction of
\cite{jarvinen05} shows a primary spot at phase 0.931, and a secondary spot 
at phase 1.325.  Due to missing phase coverage it is not possible to
verify the positioning of the primary spot from our Doppler images,
but we show that there are no significant spot features present at
phase 1.325.

The surface brightness image for November 1993
is shown in Figure~\ref{smn93}.  These images were
previously reconstructed using the the combined lines of Ca I
643.9\,nm, Fe I 666.3\,nm and Ca I 671.8\,nm by \cite{unruh95doppler}.
The images reconstructed in this work have more intermediate and low
latitude spot features.  However, there are comparable spot features
at phases 0.38 to 0.55, 0.62, 0.8 and a similar spot grouping at phase
0, though as indicated by the tick marks above the plot, there is no
phase coverage in this region.  Similar to the comparisons of the 1992
data sets, at high and polar latitudes the spot structure is weaker
and more fragmented in the images reconstructed by
\cite{unruh95doppler}.  The surface image of \cite{jarvinen05}
at epoch 1993.89 shows a primary spot at phase 1.542 and a secondary
spot at phase 1.097.  At phase 1.097 in Figure~\ref{smn93} there is a
large grouping of spots, while at phase 1.542 there is a weak spot
feature that is insignificant in strength compared to other
reconstructed spots at phases 0.38, 0.49 and 0.62.

The reconstructed surface-brightness image for November 1994 is shown
in Figure~\ref{smn94}. 
LSD has previously been applied to the 1994 CTIO data by
\cite{cameron99}.  However, in our image reconstructions the 
polar cap is stronger and less fragmented and at
mid to low latitudes the spot features are more resolved.
Examples of similar spot features include those at phase 0.1 to 0.18,
0.25 to 0.35 and 0.45 to 0.55.  For epoch 1994.87 the surface image of 
\cite{jarvinen05} shows a primary spot at phase 1.514, and a secondary spot
at phase 1.097.  While there is a spot feature at phase 1.514, it is
significantly weaker than its neighbouring features, and the spot
feature at phase 0.3 in Figure~\ref{smn94}.  We have reconstructed a
weak spot feature at the phase 1.097, but there are other stronger spots 
close by at phase 0.3.

At certain phases of the LSD profiles, it is possible to distinguish a
shallow absorption feature migrating through the line profile, which
is not reproduced in the model Maximum Entropy profiles.  Examples of
such features are shown in part II of the November 1993 data set
(Figure ~\ref{fn93b}) at phases 0.565 to 0.815.  We attribute these
absorption features as being small regions on the stellar surface that
are brighter than the surrounding photosphere in contrast to cooler
regions that produce emission features in the line profiles.  As the
imaging code is designed only to reconstruct cool features on the
stellar surface, it is not possible to reconstruct these bright
features.  The presence of these bright features does not interfere
with the reliable reconstruction of cool spots.

\section{Surface Differential Rotation}

We used the sheared-image method of Donati et al (2000) to measure the
differential rotation of AB Dor. AB Dor is an ideal candidate to
measure differential rotation as its short rotation period (0.51479-d
or 12.2053 rad d$^{-1}$) means that it is possible to observe up to
two-thirds of the stellar surface in one night and to get the
necessary overlapping phase coverage within a few days.

The image reconstruction process also incorporates a model of the stellar
surface whose rotation rate $\Omega$ depends on latitude according to
the simplified solar-like differential rotation law:

\begin{equation}
\Omega (\theta) = \Omega_{eq}\,-\,\delta\Omega cos^2 \theta     
\label{E-diffrtn}
\end{equation}

where $\Omega(\theta)$ is the rotation rate at colatitude $\theta$,
$\Omega_{eq}$ is the equatorial rotation rate and $\delta\Omega$ is
the difference between polar and equatorial rotation rates.  We
performed a large set of of image reconstructions, using a
two-dimensional grid of values for the parameters
{$\Omega_{eq}$,$\delta\Omega$}.  For each set of model parameters, the
image reconstruction was driven until it reached a fixed value of spot
filling factor. The $\chi^2$ values of the resulting images form a
"landscape" on this grid.  The best fitting model will corresponds to
the minimum in the $\chi^2$ landscape, as models with the wrong shear
will give poor fits to the data \citep{petit02hr1099}.

\begin{table*}
\fontsize{10}{14}\selectfont
\begin{tabular}{l c c c c c}
\hline
\hline

Epoch & $\Omega_{eq}$ &  d$\Omega$ & cos$^2\theta_s$ & $\Omega_s$ & n \\

&  (mrad d$^{-1}$) &  (mrad d$^{-1}$) & & (rad d$^{-1}$) \\

\hline

1992.05 & 12\,238.3 $\pm$ 2.1 & 60.1 $\pm$ 5.5 & 0.319 & 12.219 & 37761 \\
1993.89 & 12\,249.5 $\pm$ 3.5 & 71.1 $\pm$ 9.3 & 0.338 & 12.226 & 26645 \\ 
1994.87 & 12\,243.1 $\pm$ 3.1 & 73.6 $\pm$ 9.2 & 0.300 & 12.221 & 17141 \\

\hline
\hline

\end{tabular}
\caption{Summary of differential rotation parameters measured for AB Dor
at each epoch, where $\Omega_{eq}$ is the derived equatorial rotation
rate at the 1$\sigma$ 68\% confidence interval, d$\Omega$ is the
equator:pole differential rotation rate, column 4 is the inverse slope
of the ellipsoid in the $\Omega_{eq}$-$\Omega$ plane (equal to
cos$^2\theta_s$ ref. \protect\cite{donati03}), $\Omega_s$ is the
rotation rate at colatitude $\theta_s$, and n is the total number of
data points used in the imaging process.}
\label{t-diffrot}
\end{table*}

The optimally fitting differential rotation parameters and their
errors are determined by fitting the reduced $\chi^2$ landscape grids with 
a bi-dimensional paraboloid with linear and quadratic terms given by:

\begin{equation}
a \Omega_{eq}^2 + b \Omega_{eq} d\Omega + c d\Omega^2 + d \Omega_{eq} + e d\Omega
\label{e-para}
\end{equation}

The five coefficients are then used to solve for $\Omega_{eq}$ and $d\Omega$, 
and their errors.  As discussed by Donati et al. (2003), the errors are 
determined by computing the curvature radii of the $\chi^2$ paraboloid
at its minimum and the correlation coefficient parameter between the two 
differential rotation parameters.

\subsection{Results}

\begin{figure}
\psfig{file=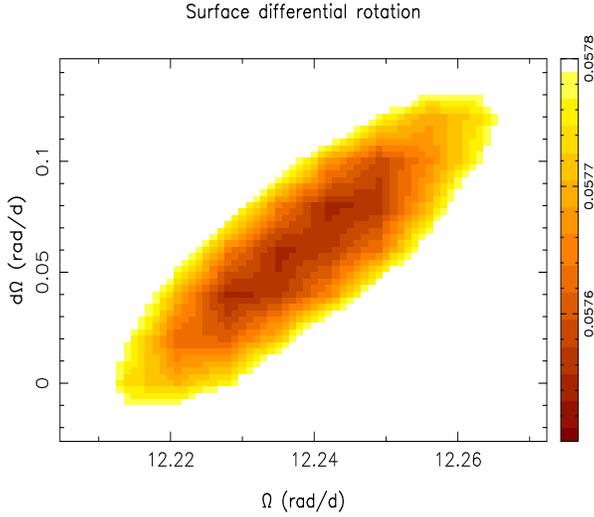,angle=270,width=7.8cm,height=6.8cm}
\caption{Reduced $\chi^2$ maps showing the  differential rotation profile 
for epoch 1992.05.  Only points that are within 1\% of the minimum
$\chi^2$ are shown. The central black region corresponds to points
where the minimum value of differential rotation was obtained, while
the outer grey points correspond to the 1.5 $\sigma$ confidence
ellipse (taking each parameter separately) on the differential
rotation parameter plain. }
\label{f-difrotel}
\end{figure}

For each epoch that comprises more than one night, we converged the
imaging code to the fixed spot filling factors obtained when
reconstructing the surface brightness images and
computed a grid of $\Omega$ and $d\Omega$ models.  The resulting
$\chi^2$ landscape is shown in Figure~\ref{f-difrotel} for epoch
1992.05 and fitting the paraboloid (equation~\ref{e-para}) to the data
resulted in the values shown in Table~\ref{t-diffrot}.  The results
for epoch 1988.96 resulted in a map without any minimum, which is
attributed to the poor quality of this data set.  The temporal
evolution of differential rotation over the epochs of this analysis is
plotted in Figure~\ref{f-difrotelall}. This plot clearly shows that
the magnitude of temporal evolution is greater than the error bars of
the differential rotation measurements, with the most noticeable change
in $d\Omega$ of 11.0 mrad d$^{-1}$ being between epochs 1992.05 and
1993.89.

\section{Discussion}

\subsection{Starspot distributions}

AB Dor is one of the most extensively Doppler imaged stars.  In this
analysis we have extended the work of  
\cite{donati97doppler,donati99doppler} and \cite{donati03} to include 
epochs 1988.96, 1992.05, 1992.95, 1993.89 and 1994.87 to have a
complete data set that has been processed using the same processing
methods.  One outstanding feature present at all epochs is a
long-lived polar cap that is slightly fragmented at epoch 1988.96 and
weaker at epoch 1992.95.  Direct evidence for the presence of a polar
cap has been shown by \cite{jeffers05pc} and \cite{jeffers06em} using
data from the Hubble Space Telescope of the RS CVn binary SV Cam.  The
reconstructed surface images of this analysis also show many small mid
to low latitude spots and a spot coverage that ranges from 6.25\% at
epoch 1992.95 to 9.2\% at epoch 1994.87.  These values bear no
resemblance to the results of long-term photometric measurements,
which show a brightness minimum in 1988 ~\citep{amado01}.  The
difference between the two results can be accounted for by the
presence of chromopheric emission, which for young stars such as AB
Dor, shows an anti-correlation with the star's photometric variations
\citep{radick98}.

For 1988.96 the reconstructed weak and fragmented polar cap is not
conclusive due to the poor signal-to-noise of the data.  However,
there is additional evidence from the results of \cite{kuerster94}
that AB Dor did not posses a polar cap in 1989.  This could show that
these large polar caps could disappear periodically.  If such
behaviour is cyclic then this could be related to a change in the
dynamo nature of the star.  The weaker polar cap for epoch 1992.95
probably results from the lower overall spot coverage at this epoch.

\begin{figure}
\psfig{file=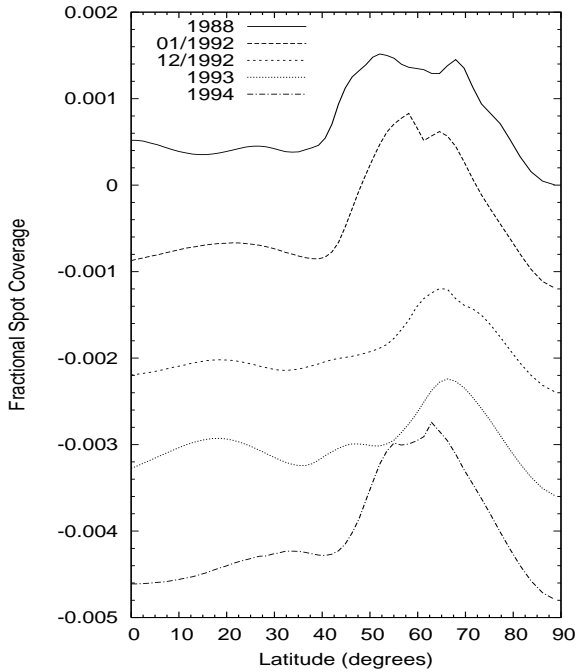,angle=0,width=8cm,height=9cm}
\caption{Variation of the fractional spot coverage per
latitude bin over the epochs of this analysis.}
\label{f-lat-tmp}
\end{figure}

The long-term stability of high, mid and low latitudes is shown to
vary on a timescale that is shorter than the temporal spacing of our
observations.  The evolution of small-scale magnetic features has been
shown by ~\cite{barnes01aper} to be on a time-scale of less than one
month for the active G2 dwarf He 699.  The fractional spot coverage
per latitude bin for each epoch of this analysis is shown in
Figure~\ref{f-lat-tmp}.  The relative fractional spot coverage for
low-latitude features (between 0$^\circ$ and 50$^\circ$) is 24\% for
1988.96, 17\% for 1992.05, 26\% for 1992.95, 30\% for 1993.89 and 19\%
for 1994.87.  The yearly distributions as shown in
Figure~\ref{f-lat-tmp} show that the peak does not become broader with
time indicating that there is no apparent migration of high latitude
spots towards the equator.  However this plot does show a global
evolution of the spot distribution with possibly the polar spot
becoming weaker with time and more spots forming at lower latitudes
(compare plots for 1993.89 and 1998.96/1992.05).  In contrast to these
results, \cite{jarvinen05} show, from photometric data of AB Dor at
similar epochs, that the mean spot latitude of 47-51$^\circ$ for
1992.05 and 45$^\circ$ for 1993.89 is significantly lower than the
values of this analysis.

\begin{figure}
\psfig{file=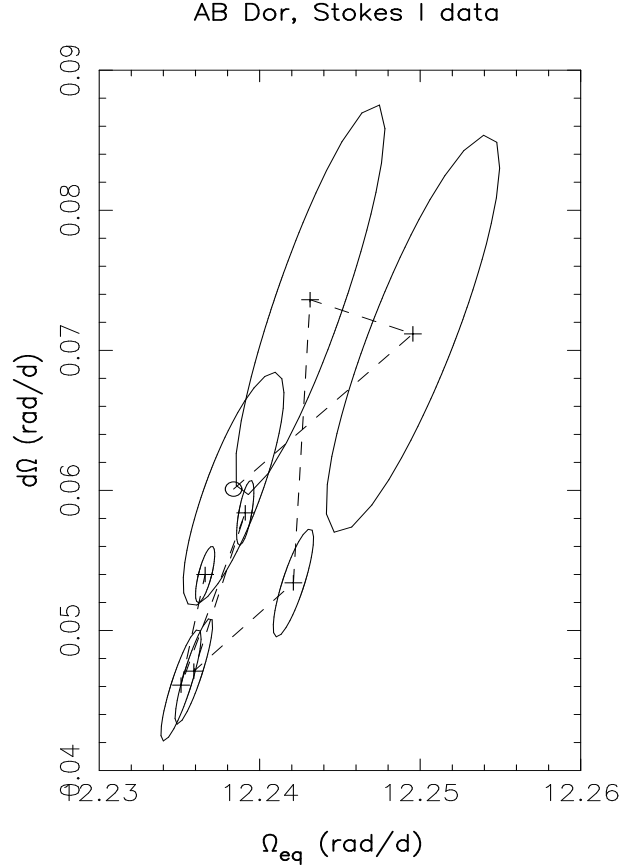,angle=0,width=8.1cm,height=11.5cm}
\caption{Differential rotation parameters measured for 
AB Dor as tabulated in Table~\ref{t-diffrot} for epochs 1992.05,
1993.89 and 1994.87 (from this work) and 1995.94, 1996.99, 1999.00,
2000.93, 2001.99 (from Donati et al. 2003).  The central point that is
indicated by $\circ$ is for epoch 1992.05 from which a dashed line
connects the points in chronological order.  The central point is the
best-fitting differential rotation measurement.  For each differential
rotation measurement, the 68\% confidence ellipse is also plotted. }
\label{f-difrotelall}
\end{figure}

The integrated brightness distributions are plotted as a function of
longitude for each surface brightness image, as shown in the lower
panel of Figures~\ref{smd88},~\ref{smj92},~\ref{smd92},~\ref{smn93},
and~\ref{smn94}.  These distributions show a variation of spot
coverage with longitude, and regions of concentrated spot coverage
that could be indicative of regions of enhanced magnetic activity or
'active longitudes'.  Active longitudes are shown to be present on AB
Dor by \cite{jarvinen05}, where they reconstruct surface images
comprising a primary and secondary spots from photometric data.
The effect of high spot coverage can in, certain cases, result in 
spurious `active longitudes' though these are always located at the 
quadrature points \citep{jeffersfs05}.

As previously discussed we show that the Doppler images reconstructed
here are in broad agreement with the location of the primary spot of
\cite{jarvinen05}, but not with the existence or location of a
secondary spot.  The difference between the results of this analysis
and those of \cite{jarvinen05} is not a result of inaccurate
photometric data, but due to that it is not possible to correlate
phases of maximum and minimum photometric brightness with regions of
the highest and lowest density of low latitude features.  This is
because the shape of the lightcurve can also be strongly influenced by
the presence of high-latitude features and an off-centred polar cap.
The effect of including high latitude spot features and polar cap is
shown in the plots of the fractional spot coverage as a function of
longitude (lower plot of
Figures~\ref{smd88},~\ref{smj92},~\ref{smd92},~\ref{smn93},
and~\ref{smn94}) where plots are shown for spot features integrated
over all latitudes and between 0$^\circ$ and 50$^\circ$.  For epochs
1988.96 and 1993.89 there is little difference in the general shape of
the two integrated spot coverage distributions.  However, for epoch
1992.05 there are notable differences at phases 0.27 to 0.4 and 0.8 to
0.95 : in particular the degree of fractional spottedness of the plot
integrated over all latitudes (approximately 0.02 for phases 0.27 to
0.4 and 0.06 for phases 0.8 to 0.95), both show the same level of
spottedness when integrated only between latitudes 0$^\circ$ to
50$^\circ$ when this is clearly not the case when integrated between
0$^\circ$ and 90$^\circ$.  The same traits can be seen in the
fractional spot distributions of epochs 1992.95 and 1994.87.  Similar
conclusions have also been reached by \cite{donati03} and
\cite{vogt99hr1099}.  Additionally, our image reconstructions also do
not show evidence for the migration of both primary and secondary
spots at a constant separation or `flip-flops' as shown in the work of
\cite{jarvinen05}.  

\subsection{Differential rotation}

Surface differential rotation was measured for epochs with sufficient
overlapping phase coverage, e.g. 1992.05, 1993.89 and 1994.87.  In a
complimentary paper, \cite{donati03diffrot} measure differential
rotation on AB Dor for epochs after 1995 using the same method of this
analysis.  The resulting values of d$\Omega$ and $\Omega_{eq}$ for all
epochs are shown in Figure~\ref{f-difrotelall}.  This plot clearly
shows that the results of this paper are generally higher than
previous results with 1994.87 having the highest differential rotation
ever measured on AB Dor.  The smaller size of the error ellipses shown
in Figure~\ref{f-difrotelall} for epochs after 1994.87 are because the
data set was taken over a longer time period enabling a more accurate
differential rotation measurement to be made.  The results so far,
from 1992.05 to 2001.99, show no evidence for any cyclic behaviour.

Another single dwarf with a similar spectral type to AB Dor and for
which there have been multiple measurements of differential rotation
is the K2V dwarf LQ Hya.  Only two measurements, a year apart, have
been made with more than a magnitude of variation in $\Delta\Omega$
ranging from 0.1942 rad d$^{-1}$ to 0.01440 rad d$^{-1}$.  The large
difference in measurements could result from the rotational period of
LQ Hya being about three times that of AB Dor and hence the spatial
resolution at the stellar surface is much less than that on AB Dor.
This could result in the measurement of differential rotation being
influenced by the short-term evolution of unresolved spots.

The temporal evolution of differential rotation of AB Dor has also
been determined by ~\cite{cameron02twist}, using a method that tracks
individual starspots in the dynamic spectrum, for the same epochs of
this analysis.  The results of this method show a stronger
differential rotation measurement for all epochs; 1992.05
($\Omega_{eq}$=12.2514 $\pm$ 0.0029 rad d$^{-1}$ and $d\Omega$=91.05
$\pm$ 13.19 mrad d$^{-1}$), 1993.97 ($\Omega_{eq}$=12.2502 $\pm$
0.0024 rad d$^{-1}$ and $d\Omega$=88.49 $\pm$ 7.47 mrad d$^{-1}$) and
1994.87 ($\Omega_{eq}$=12.2481 $\pm$ 0.0047 rad d$^{-1}$ and
$d\Omega$=66.84 $\pm$ 14.93 mrad d$^{-1}$).  \cite{donati03diffrot}
discusses the discrepancy of measurements using the two methods.  It
is concluded that there are two contributing factors; (i) the value of
$v$sin$i$ used.  The spot tracking method, which is sensitive to
$v$sin$i$, uses a value of 89\,km\,$s^{-1}$, while we use
91\,km\,$s^{-1}$ in this analysis, and (ii) the weighting of
individual spots.  The sheared imaged method of this analysis places a
higher weighting on larger spots, while the spot tracking method
places equal weighting on all spots.  However, despite these small
scale differences, the general trend of higher values of differential
rotation for the epochs of this analysis are in agreement.

Additionally, \cite{donati03diffrot} also measure differential
rotation using magnetic features, which give a different result than
using cool spots alone.  This is interpreted as being evidence that
the dynamo is distributed throughout the convective zone and not
confined at the base.  They also compare their results with models of
the differential rotation in the convection zone and show that the
internal rotation velocity field is not like that of the Sun, but more
like that of rapid rotators where the angular velocity is constant
along cylinders aligned with the rotation axis. \cite{donati03diffrot}
surmise that changes in differential rotation could result from
underlying dynamo processes.

The temporal evolution of differential rotation will also have
important consequences for the stellar structure.  Large variations
will alter the spherical oblateness of the star, such that if AB Dor
was in a close binary system it would produce long-term changes in the
star's orbital period.  This is not applicable to the AB Dor A/C
system given their comparatively large (2.3 AU) separation.  Further
support is given to the conjecture of \cite{donati03diffrot} by
~\cite{applegate92}, ~\cite{lanza98}, ~\cite{lanza99} and
~\cite{lanza05}, where theoretical interpretations of the orbital
period modulation in RS CVns is related to the operation of a
hydromagnetic dynamo in the magnetically active star.  The model of
~\cite{applegate92} assumes that these periodic modulations are caused
by the stellar magnetic cycle converting kinetic energy in the
convective zone into large-scale magnetic fields.  This is extended by
~\cite{lanza98}, and ~\cite{lanza99} to include the effect of magnetic
fields on the hydrostatic equilibrium of the magnetically active
component.  These models have been further extended by ~\cite{lanza05}
to include an improved treatment of angular momentum transport in the
stellar convective zone, but they conclude that the method of
~\cite{applegate92} is not sufficient to fully explain the mechanisms
of orbital migration in close binaries.

\section{Conclusions}

In this paper, Doppler images of the magnetically active star AB Dor
show that its starspot coverage is dominated by a long-lived and
stable polar cap and variable high to low latitude spot coverage.  The
exception to this is the surface brightness reconstruction of epoch
1988.96 where there is evidence of a weak and fragmented polar cap.
There is no cyclic behaviour found in either the latitude distribution
of spots or the spot coverage fractions.  Our surface brightness
reconstructions generally show longitudes where there is a
concentration of spots.  However, we do not find a second or minor
spot concentration which would verify the presence of active
longitudes or flip-flop cycles on AB Dor.

We have made the first measurements of differential rotation on AB Dor
for epochs 1992.05, 1993.89 and 1994.87.  The results show a temporal
evolution, with epoch 1994.87 showing the highest value of
differential rotation ever measured on AB Dor.  To a first order
approximation the temporal evolution of differential rotation show the
same variation as the results for the same data reconstructed by
\cite{cameron02twist}.  The results of this work when combined with
other previously published papers represents the first long-term
analysis of the detailed temporal evolution of differential rotation
on magnetically active stars.

\section*{Acknowledgements}

This paper is based on observations made using the 3.9\,m
Anglo-Australian Telescope, the 3.6\,m telescope at ESO and the 4\,m
telescope at CTIO.  We thank the referee Steve Saar for suggesting
several improvements to the paper.  SVJ currently acknowledges support
from a personal Marie Curie Intra-European fellowship funded within
the 6th European Community Framework Programme.  While at St Andrews
University SVJ was supported by PPARC and a scholarship from the
University of St Andrews, and would like to thank the Scottish
International Education Trust for financing a Travel Grant for a
collaborative visit to OMP.

\bibliographystyle{mn2e}
\bibliography{iau_journals,master,ownrefs}

\end{document}